\documentclass[twocolumn]{aastex631}

\usepackage{natbib}
\providecommand{\e}[1]{\ensuremath{\times 10^{#1}}}	            
\providecommand{\HII}{\ion{H}{2}}           		        	
\providecommand{\HI}{\ion{H}{1}}               		        	
\providecommand{\HA}{H$\alpha$}		                			
\providecommand{\HB}{H$\beta$}	                				
\providecommand{\abun}{12 + $\log$(O/H)}	            			

\received{16 February 2024}
\revised{25 April 2024}
\accepted{7 May 2024}
\submitjournal{AJ}

\shorttitle{Dusty Sources in I~Zw~18 with JWST}
\shortauthors{Hirschauer et al.}
\graphicspath{{./}{figures/}}

\begin{document}

\title{Imaging of I~Zw~18 by JWST.\ I.\ Detecting Dusty Stellar Populations}

\correspondingauthor{Alec S.\ Hirschauer}
\email{ahirschauer@stsci.edu}

\author[0000-0002-2954-8622]{Alec S.\ Hirschauer}
\affil{Space Telescope Science Institute, 3700 San Martin Drive, Baltimore, MD 21218, USA}

\author[0000-0001-7866-8738]{Nicolas Crouzet}
\affil{Leiden Observatory, Leiden University, P.O.\ Box 9513, 2300 RA Leiden, The Netherlands}

\author[0000-0002-2667-1676]{Nolan Habel}
\affil{Jet Propulsion Laboratory, California Institute of Technology, 4800 Oak Grove Dr., Pasadena, CA 91109, USA}

\author[0000-0003-4023-8657]{Laura Lenki\'{c}}
\affil{Stratospheric Observatory for Infrared Astronomy, NASA Ames Research Center, Mail Stop 204-14, Moffett Field, CA 94035, USA}
\affil{Jet Propulsion Laboratory, California Institute of Technology, 4800 Oak Grove Dr., Pasadena, CA 91109, USA}

\author[0000-0002-7512-1662]{Conor Nally}
\affil{Institute for Astronomy, University of Edinburgh, Blackford Hill, Edinburgh, EH9 3HJ, UK}

\author[0000-0003-4870-5547]{Olivia C.\ Jones}
\affil{UK Astronomy Technology Centre, Royal Observatory, Blackford Hill, Edinburgh, EH9 3HJ, UK}

\author[0009-0003-6182-8928]{Giacomo Bortolini}
\affil{The Oskar Klein Centre, Department of Astronomy, Stockholm University, AlbaNova, SE-10691 Stockholm, Sweden}

\author[0000-0003-4850-9589]{Martha L.\ Boyer}
\affil{Space Telescope Science Institute, 3700 San Martin Drive, Baltimore, MD 21218, USA}

\author[0000-0003-1689-9201]{Kay Justtanont}
\affil{Chalmers University of Technology, Dept.\ Space, Earth and Environment, Onsala Space Observatory, 439 92 Onsala, Sweden}

\author[0000-0002-0522-3743]{Margaret Meixner}
\affil{Jet Propulsion Laboratory, California Institute of Technology, 4800 Oak Grove Dr., Pasadena, CA 91109, USA}

\author[0000-0002-3005-1349]{G{\"o}ran {\"O}stlin}
\affil{The Oskar Klein Centre, Department of Astronomy, Stockholm University, AlbaNova, SE-10691 Stockholm, Sweden}

\author[0000-0001-7416-7936]{Gillian S.\ Wright}
\affil{UK Astronomy Technology Centre, Royal Observatory, Blackford Hill, Edinburgh, EH9 3HJ, UK}


\author[0000-0002-0438-0886]{Ruyman Azzollini}
\affil{Mullard Space Science Laboratory, University College London, Holmbury St Mary, Dorking, Surrey, RH5 6NT, UK}

\author[0000-0002-5797-2439]{Joris A.\ D.\ L.\ Blommaert}
\affil{Astronomy and Astrophysics Research Group, Department of Physics and Astrophysics, VU Brussel, Pleinlaan 2, B-1050 Brussels, Belgium}

\author[0000-0001-9737-169X]{Bernhard Brandl}
\affil{Leiden Observatory, Leiden University, P.O.\ Box 9513, 2300 RA Leiden, The Netherlands}

\author[0000-0002-5342-8612]{Leen Decin}
\affil{Instituut voor Sterrenkunde, KU Leuven, Celestijnenlaan 200D, 3001 Leuven, Belgium}


\author[0000-0001-6576-6339]{Omnarayani Nayak}
\affil{Space Telescope Science Institute, 3700 San Martin Drive, Baltimore, MD 21218, USA}
\affil{NASA Goddard Space Flight Center, 8800 Greenbelt Road, Greenbelt, MD 20771, USA}

\author[0000-0001-9341-2546]{Pierre Royer}
\affil{Instituut voor Sterrenkunde, KU Leuven, Celestijnenlaan 200D, 3001 Leuven, Belgium}

\author[0000-0001-9855-8261]{B.\ A.\ Sargent}
\affil{Space Telescope Science Institute, 3700 San Martin Drive, Baltimore, MD 21218, USA}
\affil{Department of Physics \& Astronomy, Johns Hopkins University, 3400 N.\ Charles St., Baltimore, MD 21218, USA}


\author[0000-0001-5434-5942]{Paul van der Werf}
\affil{Leiden Observatory, Leiden University, P.O.\ Box 9513, 2300 RA Leiden, The Netherlands}




\begin{abstract}

\noindent 
We present a JWST imaging survey of I~Zw~18, the archetypal extremely metal-poor, star-forming (SF), blue compact dwarf galaxy.
With an oxygen abundance of only $\sim$3\% $Z_{\odot}$, it is among the lowest-metallicity systems known in the local Universe, and is, therefore, an excellent accessible analog for the galactic building blocks which existed at early epochs of ionization and star formation.
These JWST data provide a comprehensive infrared (IR) view of I~Zw~18 with eight filters utilizing both Near Infrared Camera (F115W, F200W, F356W, and F444W) and Mid-Infrared Instrument (F770W, F1000W, F1500W, and F1800W) photometry, which we have used to identify key stellar populations that are bright in the near- and mid-IR.
These data allow for a better understanding of the origins of dust and dust-production mechanisms in metal-poor environments by characterizing the population of massive, evolved stars in the red supergiant (RSG) and asymptotic giant branch (AGB) phases.
In addition, it enables the identification of the brightest dust-enshrouded young stellar objects (YSOs), which provide insight into the formation of massive stars at extremely low metallicities typical of the very early Universe.
This paper provides an overview of the observational strategy and data processing, and presents first science results, including identifications of dusty AGB, RSG, and bright YSO candidates.
These first results assess the scientific quality of JWST data and provide a guide for obtaining and interpreting future observations of the dusty and evolved stars inhabiting compact dwarf SF galaxies in the local Universe.

\end{abstract}




\section{Introduction} 
\label{sec:intro}

Metal-poor star-forming (SF) dwarf galaxies in the local Universe represent accessible analogs to those at high redshift (e.g., \citealp{bib:MotinoFlores2021}).
As the enrichment history for a given system traces the buildup of heavy elements from successive generations of stellar nucleosynthesis, a low-abundance galaxy mimics the astrophysical conditions common in the early Universe, including the universal epoch of peak star formation (``cosmic noon," $z$$\sim$ 1.5--2; \citealp{bib:Madau2014, bib:VanSistine2016}), at which point a majority of the Universe's star formation and chemical enrichment is expected to have taken place.
At the very lowest metallicities, we may therefore approximate the SF environments of the time period shortly after the Big Bang.
I~Zw~18 is among the most extremely metal-poor (XMP; $Z$$<$ 0.1 $Z_{\odot}$) systems known (\abun =7.17 $\pm$ 0.04), with a measured gas-phase oxygen abundance of only approximately 3\% solar (e.g., \citealp{bib:Alloin1979, bib:Lequeux1979, bib:French1980, bib:KinmanDavidson1981, bib:Pagel1992, bib:SkillmanKennicutt1993, bib:Izotov1999, bib:Lebouteiller2013}).
At a distance of 18.2 $\pm$ 1.5 Mpc \citep{bib:Aloisi2007} and with global star-formation rate (SFR) values measured at $\sim$0.13 $M_{\odot}$ yr$^{-1}$ (computed via \HI\ absorption; \citealp{bib:Lebouteiller2013}) to $\sim$0.17 $M_{\odot}$ yr$^{-1}$ (estimated from radio free-free continuum emission; \citealp{bib:Hunt2005a}), with localized values as high as $\sim$1 $M_{\odot}$ yr$^{-1}$ (via statistical analysis; \citealp{bib:Annibali2013}) it is an ideal laboratory for study of both the young and evolved star demographics in an environment analogous to that found in the very early Universe.
Typical of such systems, I~Zw~18 has an enhanced total X-ray luminosity per unit of SFR \citep{bib:Lebouteiller2017}.
Additional properties of I~Zw~18 are listed in Table \ref{tab:properties}.

As the archetypal blue compact dwarf (BCD) galaxy, I~Zw~18 is currently experiencing a period of intensely elevated SFR \citep{bib:JanowieckiSalzer2014, bib:Janowiecki2017} and has thus been the subject of focused observations since its discovery \citep{bib:Zwicky1966, bib:SargentSearle1970, bib:SearleSargent1972}.
Early descriptions of this system identified two bright regions of star formation separated by 5\arcsec.8 which constitute the main body, subsequently referred to as the northwest (NW) and southeast (SE) components.
Active star formation in the brighter NW component was established via observations tracing ionized gas emission (e.g., \citealp{bib:HunterThronson1995}) and Wolf--Rayet (WR) stars (e.g., \citealp{bib:Izotov1997, bib:Legrand1997, bib:deMello1998, bib:Brown2002}).
This region has an \HI\ envelope which contains inherently metal-free gas pockets \citep{bib:Lebouteiller2013} amongst inhomogeneities of the multiphase interstellar medium (ISM; \citealp{bib:James2014}).
An additional diffuse SF region $\sim$22\arcsec\ NW of the main body (known as Component C) was identified later \citep{bib:Davidson1989, bib:DufourHester1990}, with radial velocity studies of the ionized and neutral (\HI) gas demonstrating its physical association and common \HI\ envelope with I~Zw~18, respectively (e.g., \citealp{bib:Dufour1996, bib:IzotovThuan1998, bib:vanZee1998}).

I~Zw~18 has been intensely targeted by the Hubble Space Telescope (HST), and while initial studies found no evidence for an old, existing stellar population (e.g., \citealp{bib:HunterThronson1995}), subsequent work utilizing deeper data confirmed the presence of an early star-formation episode based on evidence of asymptotic giant branch (AGB) stars \citep{bib:Aloisi1999} and gas-phase abundance study \citep{bib:Garnett1997}.
Details concerning an intense, recent bout of star formation which began at least $\sim$30--50 Myr ago were published by \citet{bib:Dufour1996}, with the near-infrared (near-IR) study of \citet{bib:Ostlin2000} concluding that the stellar population of I~Zw~18 is made up of a younger population of red supergiants (RSGs) and an older population of AGB stars, also found by \citet{bib:IzotovThuan2004}.
Subsequent study of these evolved stars by \citet{bib:OstlinMouhcine2005} employed HST medium-band filters to additionally identify a subset of five AGB stars via C$_{2}$ absorption.
The underlying stellar population of Component C was found to be similarly old, while exhibiting some moderate recent activity \citep{bib:Aloisi1999, bib:IzotovThuan2004, bib:Jamet2010}.

\begin{deluxetable*}{lcc} 
\label{tab:properties}
\tabletypesize{\small}
\tablewidth{0pt}
\tablecaption{
Properties of I~Zw~18
}
\tablehead{
\colhead{Observed Global Quantities}&\colhead{Value}&\colhead{Reference}
}
\startdata
R.A.\ (J2000) & 09:34:02 & \citet{bib:IzotovThuan2016} \\
Decl.\ (J2000) & +55:14:28 & \citet{bib:IzotovThuan2016} \\
Distance & 18.2 $\pm$ 1.5 Mpc & \citet{bib:Aloisi2007} \\
Distance modulus & 31.30 $\pm$ 0.17 & \citet{bib:Aloisi2007} \\
Recessional velocity & 751 $\pm$ 2 km s$^{-1}$ & \citet{bib:Thuan1999} \\
Redshift & 0.00251 & \citet{bib:IzotovThuan2016} \\
Stellar mass ($M_{*}$) & 2.0 $\times$ 10$^{7}$ M$_{\odot}$ & \citet{bib:Madden2014} \\
Gas mass (\HI) & 1.0 $\times$ 10$^{8}$ M$_{\odot}$ & \citet{bib:Lelli2012} \\
Dust mass & 1.1 $\times$ 10$^{4}$ M$_{\odot}$ & \citet{bib:Herrera-Camus2012} \\
Extinction ($A_{V}$) & 0 -- 0.2 mag & \citet{bib:IzotovThuan2016} \\
SFR & $\sim$0.1-- $\sim$1 M$_{\odot}$ yr$^{-1}$ & \citet{bib:Lebouteiller2013, bib:Annibali2013} \\
SFR/$M_{*}$ (sSFR) & 10$^{-7}$--10$^{-8}$ yr$^{-1}$ & A.\ Aloisi (2024, private communication) \\
\abun\ & 7.17 $\pm$ 0.04 & \citet{bib:SkillmanKennicutt1993} \\
\emph{U -- B} & --0.88 & \citet{bib:vanZee1998} \\
\emph{B -- V} & --0.03 & \citet{bib:vanZee1998} \\
\enddata
\end{deluxetable*}

\vspace{-25pt}
Establishing a census of evolved stars in I~Zw~18 will help to provide for a better understanding of the origins and role of interstellar dust at the earliest epoch of our Universe.
While supernovae (SNe) have been found to quickly produce dust in significant quantities locally, it is unclear how similarly this mechanic behaves at the extremely low levels of metal enrichment which characterized galaxies at high redshift.
In addition, the rate of both new and existing dust destruction by SNe via the passage of shocks may rival the rate of creation \citep{bib:Temim2015}!
AGB stars are known to be a major source of dust production within nearby systems (e.g., \citealp{bib:Matsuura2009}).
As they evolve, thermal pulsations and convective cell activity push the products of nucleosynthetic reactions {from the internal regions previously exposed to helium burning reactions towards toward the stellar surface (e.g., \citealp{bib:Iben1974, bib:Iben1975, bib:Lattanzio1987, bib:Lattanzio1993, bib:BoothroydSackmann1988}), where its ejection enriches the surrounding ISM.
Because AGB stars represent the final evolutionary stage for the vast majority of stars which have left the main sequence (MS), their contributions of dust and heavy elements to the ISM due to mass loss, and the subsequent evolutionary effects on the ISM and other stars, is significant \citep{bib:Hofner2018}.

In metal-poor environments and at high redshift, the dust contribution from these AGB stars is currently not well defined.
For ancient galaxies populating the early Universe, the prevailing expectation had been that their low- to intermediate-mass MS progenitor stars would have required a greater amount of time than had presently elapsed in order to evolve to the AGB phase, suggesting that they cannot contribute significantly to the overall dust budget.
Recent work, however, has begun casting doubt on this conventional wisdom, with evidence for such evolved stars in the low-metallicity galaxy Sextans A suggesting that these stars could contribute dust at early epochs.
Additionally, observational evidence for large reservoirs of dust existing at redshifts up to $z$$\sim$ 6 has been found \citep{bib:Bertoldi2003, bib:Robson2004, bib:Beelen2006, bib:Algera2023}, and perhaps even as distant as $z$$>$ 10 \citep{bib:Curtis-Lake2023}.
Furthermore, studies of nearby low-abundance systems such as dwarf galaxies and globular clusters have shown that these environments possess dust originating from evolved stars \citep{bib:McDonald2010, bib:Whitelock2018, bib:Jones2018}; however, the effects of metallicity on AGB star dust production are disputed \citep{bib:vanLoon2005, bib:McDonald2011, bib:Sloan2012, bib:Sloan2016, bib:Dell'Agli2019b}.

\begin{figure*} 
\centering
\includegraphics[width=1.0\textwidth]{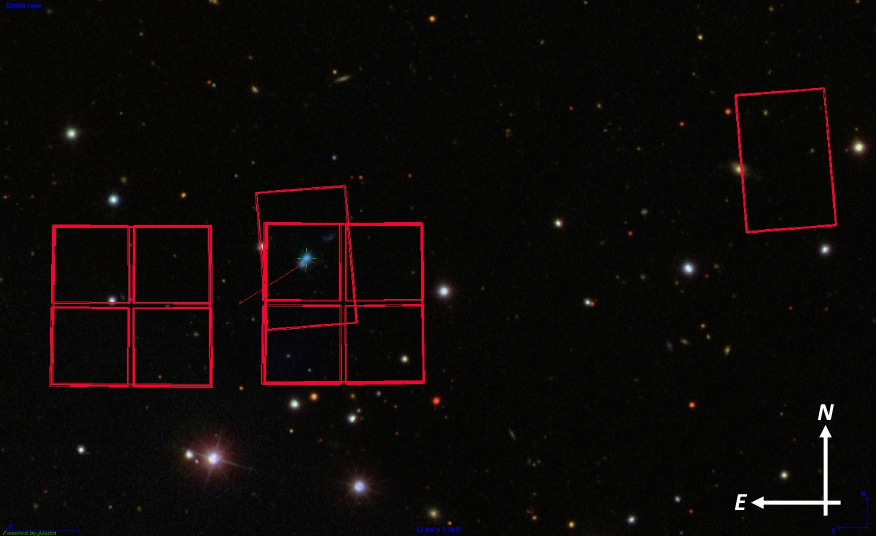}\hfill
\caption{
Imaging footprints for I~Zw~18 with both NIRCam (squares) and MIRI (rectangles) overlaid upon an optical-band image adopted from the Sloan Digital Sky Survey \citep{bib:York2000, bib:Abazajian2004}.
North is up, while east is to the left.
The compact extent of the galaxy, seen as the small, blue, double-lobed object, allows for only a single pointing with each instrument.
A small position offset (+55\arcsec.0 in \emph{x}; +35\arcsec.0 in \emph{y}) from center (between the two NIRCam modules) ensured I~Zw~18's placement within the upper-left quadrant of NIRCam Module B (B3), regardless of the spacecraft's position angle.
While NIRCam Module A (middle left) provides a suitable region of blank space for performing background correction, for MIRI a coordinated parallel observation was included (upper right).
}
\label{fig:FoVs}
\end{figure*}

Observational efforts in the IR searching for metal-poor local galaxies possessing populations of evolved stars are ongoing.
Prior to the advent of JWST, the state of the art for resolved extragalactic stellar populations was the Surveying the Agents of a Galaxy's Evolution (SAGE) program for the Large \citep{bib:Meixner2006, bib:Blum2006, bib:Whitney2008, bib:Bernard2008, bib:Jones2015b} and Small \citep{bib:Gordon2011, bib:Boyer2011, bib:Sewilo2013} Magellanic Clouds (LMC/SMC), taken with the Spitzer Space Telescope (Spitzer).
In addition to providing censuses of evolved stars, dust-production and mass-injection rates into the ISM were quantified (e.g., \citealp{bib:Matsuura2009, bib:Boyer2012, bib:Riebel2012, bib:Srinivasan2016}), with significant dust production (10$^{-4}$ $M_{\odot}$ yr$^{-1}$) via oxygen-rich AGB stars found in the LMC ($\approx$$\frac{1}{2}$ $Z_{\odot}$).
The DUST in Nearby Galaxies with Spitzer survey (DUSTiNGS; \citealp{bib:Boyer2015a, bib:Boyer2015b, bib:McQuinn2017, bib:Boyer2017, bib:Goldman2019a}) sampled 50 additional nearby (within 1.5 Mpc) galaxies using Spitzer 3.6 and 4.5 $\mu$m observations, finding that AGB stars are a major contributor of interstellar dust even at low metallicities.
Recent works observing the nearby SF galaxy NGC 6822 \citep{bib:Lenkic2024, bib:Nally2024} with the JWST \citep{bib:Gardner2023} showcase its unparalleled IR sensitivity, improving upon earlier studies utilizing Spitzer and ground-based data \citep{bib:Jones2019, bib:Hirschauer2020}.
Even with the crowded fields and large distance to I~Zw~18, JWST is capable of detecting and resolving individual evolved red stars, including RSGs and both oxygen- and carbon-rich AGB stars, as well as those undergoing substantial dust loss via superwind in which the most significant amounts of dust are produced (i.e., ``x-AGB" stars; \citealp{bib:Blum2006, bib:Boyer2011, bib:Boyer2015a}).

\begin{deluxetable*}{ccccccc} 
\label{tab:obs_params_NIRCam_MIRI}
\tabletypesize{\small}
\tablewidth{0pt}
\tablecaption{
Observing Parameters for NIRCam and MIRI Imaging
}
\tablehead{
\colhead{Filter}&\colhead{Field}&\colhead{Readout Pattern}&\colhead{Groups/Int.}&\colhead{Int./Exp.}&\colhead{Dithers}&\colhead{Total Exp.\ Time}
\\
& & & & & & [sec]
}
\startdata
F115W & NIRCam Prime & \textsc{shallow4} & 6 & 1 & 4 & 1245.465 \\
F200W & NIRCam Prime & \textsc{shallow4} & 6 & 1 & 4 & 1245.465 \\
F356W & NIRCam Prime & \textsc{shallow4} & 6 & 1 & 4 & 1245.465 \\
F444W & NIRCam Prime & \textsc{shallow4} & 6 & 1 & 4 & 1245.465 \\
\hline
F770W & MIRI Prime & \textsc{fastr1} & 100 & 1 & 4 & 1110.016 \\
F1000W & MIRI Prime & \textsc{fastr1} & 200 & 1 & 4 & 2220.032 \\
F1500W & MIRI Prime & \textsc{fastr1} & 110 & 2 & 4 & 2453.135 \\
F1800W & MIRI Prime & \textsc{fastr1} & 88 & 6 & 4 & 5916.385 \\
\hline
F1000W & MIRI Parallel & \textsc{fastr1} & 112 & 1 & 4 & 1243.218 \\
F1500W & MIRI Parallel & \textsc{fastr1} & 112 & 1 & 4 & 1243.218 \\
\enddata
\tablecomments{Observations for JWST PID:\ 1233 executed on 2022 October 27.}
\end{deluxetable*}

\vspace{-25pt}
JWST additionally offers incomparable sensitivity to dust-enshrouded young stellar objects (YSOs) and pre-main-sequence (PMS) stars (e.g., \citealp{bib:Jones2023}).
The XMP abundance level, high SFR, and striking blue appearance of I~Zw~18 affirms the extreme nature of its SF regions.
As YSOs are born in regions of active star formation, they exhibit strong IR excess as their light is absorbed and reemitted by their surrounding cool dusty envelopes and accretion disks.
Previous photometric studies with earlier facilities were only capable of detecting the most massive such stars ($\gtrsim$8$M_{\odot}$), which account for only roughly one in every 10,000, and star clusters \citep{bib:Whitney2008, bib:GruendlChu2009, bib:Carlson2012, bib:Nayak2023}.
Because YSOs rapidly progress through the different stages of their evolution, observations are relatively rare.
Detailed study of the population of young stars in I~Zw~18 will allow for a more complete understanding of the physical mechanisms which govern star formation in environments typical of the very early Universe.

For the first time, we present high-quality near- and mid-IR imaging of the BCD galaxy I~Zw~18 taken with JWST, which with its XMP level of chemical enrichment and high SFR represents an excellent accessible analog for high-redshift SF systems.
In \S\ref{sec:obsprog}, we describe the observing strategy employed for this Guaranteed Time Observational (GTO) program (PID:\ 1233;\ PI:\ M.\ Meixner), while in \S\ref{sec:data} we describe the data processing and analysis.
Section \ref{sec:earlyresults} presents our initial results, including the first images of I~Zw~18 obtained using JWST as well as characterizations of various stellar source type candidates.
We summarize the program in \S\ref{sec:summary}.


\section{Observing Program} 
\label{sec:obsprog}

We imaged I~Zw~18 with JWST utilizing both the Near Infrared Camera (NIRCam; \citealp{bib:Rieke2005, bib:Rieke2023}) and the Mid-Infrared Instrument (MIRI; \citealp{bib:Rieke2015, bib:Wright2023}) imager \citep{bib:Bouchet2015, bib:Dicken2024}.
In total, 6.49 hr of integration time were split between NIRCam ($\sim$1.91 hr) and MIRI ($\sim$4.58 hr), across eight filters (four filters each) employed for the prime observations.
The observations with NIRCam were taken on 2022 October 27, utilizing the F115W and F200W short-wavelength (SW) filters, alongside the F356W and F444W long-wavelength (LW) filters, with I~Zw~18 centered in the upper-left quadrant of Module B (B3), and Module A observing empty sky to provide suitable data for background contamination subtraction.
Employing both the A and B modules with a single pointing, these observations covered a total field of view (FOV) of 9.7 arcmin$^{2}$, with two 2\arcmin.2 $\times$ 2\arcmin.2 fields separated by a $\sim$44\arcsec\ gap.

Observations with MIRI were taken with a single pointing on the same date, covering a FOV of 74\arcsec\ $\times$ 113\arcsec.
These utilized the F770W, F1000W, F1500W, and F1800W filters, centered on the galaxy.
In addition, MIRI coordinated parallel observations were obtained of an off-target region of empty space with the F1000W and F1500W filters, also in a single pointing 74\arcsec\ $\times$ 113\arcsec\ in size.

The spatial coverage for the two instruments is illustrated in Figure \ref{fig:FoVs}, and the principal characteristics of this program are summarized in Table \ref{tab:obs_params_NIRCam_MIRI}.
At a distance of 18.2 Mpc \citep{bib:Aloisi2007}, the angular resolution is 0\arcsec.04 -- 0\arcsec.14 ($\sim$3.5 -- 12.4 pc) in the NIRCam filters and 0\arcsec.3 -- 0\arcsec.6 ($\sim$26.5 -- 53.1 pc) in the MIRI filters, respectively.
This provides an improvement over existing IR imaging with Spitzer (spatial resolution $\sim$2\arcsec) by greater than an order of magnitude, and is of similar quality to optical-band imaging with HST.

\begin{figure*} 
\centering
\includegraphics[width=1.0\textwidth]{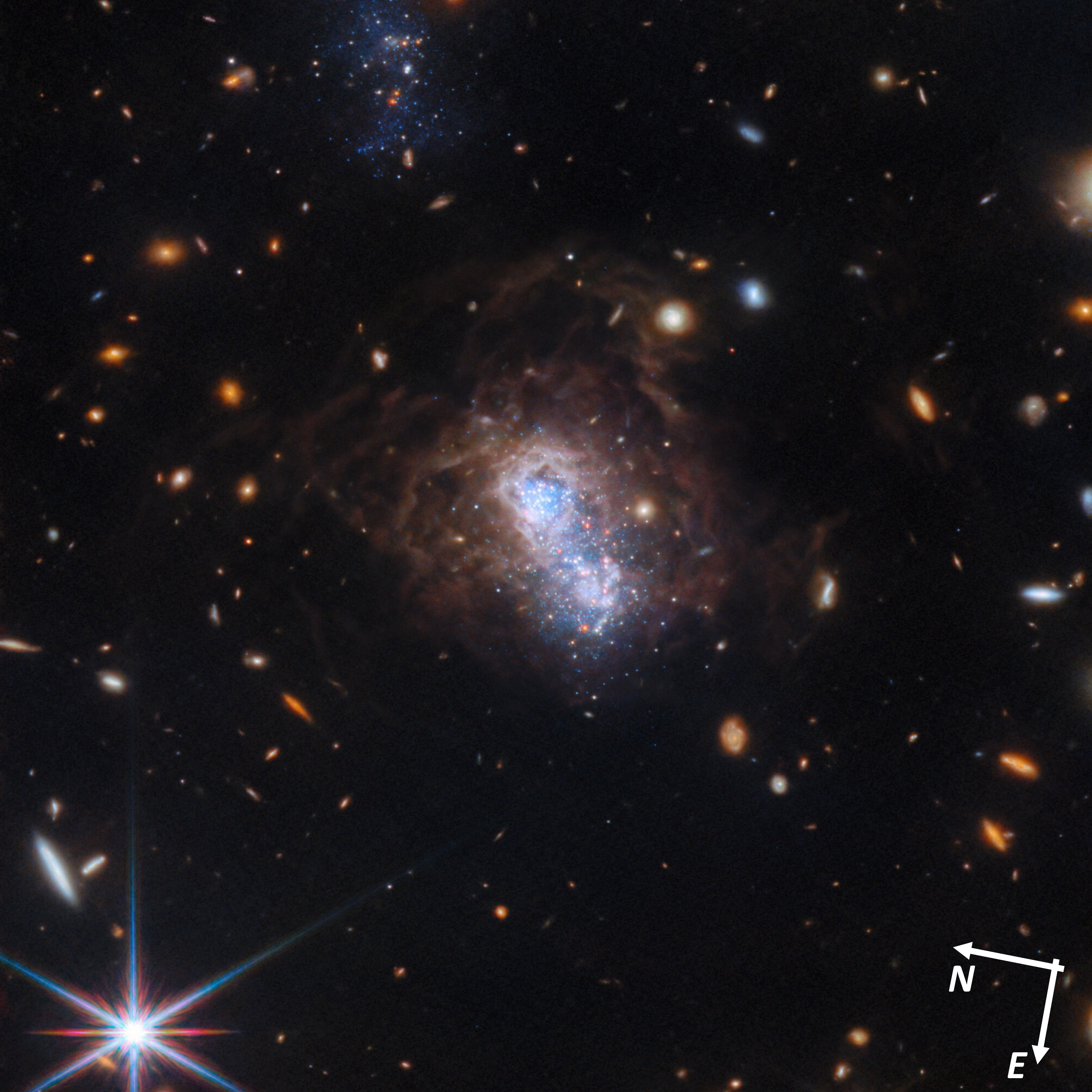}\hfill
\caption{
Four-color near-IR composite image of I~Zw~18 in NIRCam F115W (dark blue), F200W (light blue), F356W (orange), and F444W (red).
Image orientation is such that north is to the left, while east is down.
An underlying red population of older stars is complemented by hot, massive stars, identifiable by bright, blue sources occupying the NW and SE components.
These \HII\ regions are swaddled amongst a cocoon of gas and dust, while resolved background galaxies populate the image outskirts.
Stellar sources affiliated with Component C are visible at the top of the image, just left of center.
The image resolution afforded by NIRCam imaging is 0\arcsec.04 -- 0\arcsec.14, roughly equivalent to that of HST's optical bands.
This image was made by ESA/NASA/CSA/STScI.
}
\label{fig:prettypicture}
\end{figure*}

\subsection{NIRCam Strategy} 
\label{sec:NIRCam_strat}

Our NIRCam observations were taken with the \textsc{full} frame and without need for a primary dither pattern, as there were no mosaic gaps that needed to be filled in.
The \textsc{shallow4} read-out pattern was selected in effort to optimize the signal-to-noise ratio (S/N), with one integration per exposure and six groups per integration.
In addition, we implemented a four-point subpixel dither pattern to sample the point spread function (PSF), which resulted in an overall exposure time of 1245.5 s for each of the SW+LW filter combinations (i.e., F115W+F356W and F200W+F444W).

Our selection of NIRCam filters was based on prior work on stellar populations at these wavelengths.
The F115W and F200W filters are similar to standard Johnson \emph{J} and \emph{K$_{s}$}, respectively, while the F356W and F444W filters are similar to Spitzer IRAC [3.6] and [4.5], respectively.
This allows for comparative studies with previous observations made of other galaxies, including the SAGE studies of the Magellanic Clouds (e.g., \citealp{bib:Blum2006, bib:Meixner2006, bib:Whitney2008, bib:Gordon2011, bib:Boyer2011}), the DUSTiNGS project, which observed many nearby SF galaxies (e.g., \citealp{bib:Boyer2015a, bib:Boyer2015b, bib:McQuinn2017, bib:Goldman2019a}), and others (e.g., \citealp{bib:NikolaevWeinberg2000, bib:Jones2015, bib:Jones2018, bib:Goldman2019b, bib:Jones2019, bib:Hirschauer2020}).

Centering the spatial extent of I~Zw~18 within the upper-left quadrant of Module B (B3) was accomplished by requesting a special requirement which offset the position of the target by 55\arcsec.0 in the ``\emph{X}" direction and 35\arcsec.0 in the ``\emph{Y}" direction, centered at R.A.\ = 09:34:02.10, decl.\ = +55:14:25.00.
This allows for uninterrupted spatial coverage of I~Zw~18 in the NIRCam SW FOV without relying on a restriction in spacecraft position angle.

\subsection{MIRI Strategy} 
\label{sec:MIRI_strat}

Observations with MIRI employed a four-point \textsc{cycling} dither pattern and a \textsc{small} pattern size, utilizing the \textsc{full} subarray and \textsc{fastr1} read-out pattern, centered at R.A.\ = 09:34:02.00, decl.\ = +55:14:28.00.
For the F770W and 1000W filters, we implemented strategies of 100 and 200 groups per integration, respectively, with one integration per exposure, for four total integrations each.
The total exposure time for the F770W filter was 1110.0 s, while for F1000W it was 2220.0 s.
With the next-longest-wavelength filter, F1500W used 110 groups per integration and two integrations per exposure, resulting in eight total integrations totaling 2453.1 s.
The longest-wavelength MIRI filter, F1800W, necessitated selecting 88 groups per integration to avoid saturation of the sky background, achieving a long total exposure time of 5916.4 s by increasing the number of integrations per exposure to six, for 24 total integrations.

Our selection of MIRI filters was guided by the predicted JWST fluxes and color separations for IR-bright populations from \citet{bib:Jones2017}.
Color--magnitude diagrams (CMDs) utilizing these long wavelengths are sensitive to the reddest and dustiest sources, offering unambiguous identification of evolved (RSGs and AGB stars) and young (YSOs) stellar populations via long-wavelength baselines.
The longer-wavelength MIRI filters F2100W and F2550W were \emph{not} chosen despite the wider possible baselines of its CMDs, owing to a steep drop-off in sensitivity which would have required substantially longer integration times to achieve comparable S/N.
Separation of source types was determined to be equivalently achievable using the F1800W filter with shorter exposure times.
Finally, detection of polycyclic aromatic hydrocarbon (PAH) emission is accomplished with the F770W filter, and the F1000W filter is sensitive to the 10 $\mu$m silicate feature.

In addition, a MIRI coordinated parallel observation (when NIRCam was prime) was included to provide background comparison images useful for contaminant source subtraction.
The MIRI coordinated parallel image was obtained in the F1000W and F1500W filters utilizing a \textsc{full} subarray, and designed to maximize available integration time as determined by the NIRCam prime observations.
We utilized a \textsc{fastr1} read-out pattern, with 112 groups per integration and one integration per exposure over the four total dithers, resulting in a total exposure time of 1243.2 s per filter.
Data acquired from the MIRI parallel field are not considered further in this paper.


\section{Data Reduction and Source Extraction} 
\label{sec:data}

\subsection{Data-reduction Methodology} 
\label{sec:reduction}

Processing of the NIRCam imaging data was carried out utilizing a slightly modified version of the JWST official pipeline (version 1.7.2) through all three stages.
These modifications corrected for 1/\emph{f} noise and flat-field correction noise, World Coordinate System (WCS) alignment issues, and included the most recent NIRCam calibration files at the time of data processing.
For Stage-1 data, we used {\texttt{jwst\_1069.pmap} of the Operational Pipeline Calibration Reference Data System (CRDS), produced on 2022 October 3 with on-sky derived photometric zero-points \citep{bib:Boyer2022, bib:Gordon2022}.
A frame0 correction was implemented to recover stars saturated in the first group ($\sim$21 s), but were unsaturated in the first 10.7 s comprising frame0 (\texttt{ramp\_fit.suppress\_one\_group=False}).
For Stage-2 output files (*\_cal.fits), \texttt{image1overf.py}\footnote{\protect\url{https://github.com/chriswillott/jwst}} was used to correct for 1/\emph{f} noise.
Because the small FOV contained few suitable Gaia reference stars, astrometric alignment was done relatively between frames and filters.
A simple source catalog was created from a single arbitrary exposure in the F115W filter using source detection from the JWST pipeline.
Alignment of the remaining F115W frames and frames from the remaining filters to this source catalog was accomplished using the JWST/Hubble Alignment Tool (JHAT)\footnote{\protect\url{https://github.com/arminrest/jhat}}.
Because alignment to \emph{Gaia} was already accomplished via JHAT, the Stage-3 tweakreg step was skipped.

\begin{deluxetable*}{l|cccccc} 
\label{tab:starbugparams}
\tabletypesize{\small}
\tablewidth{0pt}
\tablecaption{
{\sc starbugii} Parameters Used for All Photometry
}
\tablehead{
\colhead{Parameter}&\colhead{F115W}&\colhead{F200W}&\colhead{F356W}&\colhead{F444W}&\colhead{F770W}
}
\startdata
        SIGSRC       & 5.4 & 5.5 & 4.8 & 4.8        & 6.5 \\ 
        SIGSKY       & 2 & 2.1 & 2 & 1.95           & 2.5 \\ 
        RICKER\_R    & 1 & 1 & 1 & 1                & 1 \\ 
        SHARP\_LO    & 0.4 & $-0.2$ & 0 & 0         & 0 \\ 
        SHARP\_HI    & 1.03 & 1.7 & 1.1 & 1.15      & 0.9 \\ 
        ROUND\_LO/HI & $\pm$2 & $\pm$2 & $\pm$2 & $\pm$2   & $\pm$2 \\ 
        \hline
        APPHOT\_R    & 1.5 & 1.5 & 1.5 & 1.5        & 2 \\ 
        SKY\_RIN     & 3 & 3 & 3 & 3                & 4 \\ 
        SKY\_ROUT    & 4.5 & 4.5 & 4.5 & 4.5        & 6 \\ 
        BOX\_SIZE    & 2 & 2 & 2 & 2                & 5 \\ 
        CRIT\_SEP    & 6 & 6 & 8 & 8                & 8 \\ 
        \hline
        MATCH\_THRESH & 0.06 & 0.06 & 0.1 & 0.1     & 0.1 \\ 
        NEXP\_THRESH   & 2 & 2 & 2 & 2              & 2 \\ 
\enddata
\end{deluxetable*}

\vspace{-25pt}
JWST pipeline version 1.12.0 with CRDS version 11.17.9 and context \texttt{jwst\_1149.pmap} were used in the creation of the final F770W imaging data.
The remaining MIRI filters were subsequently processed using CRDS version 11.17.6 and context \texttt{jwst\_1179.pmap}.
The new CRDS context included updates for the MIRI Medium Resolution Spectrometer and thus does not affect the MIRI imaging data we present here.
Flux-calibrated images in all four filters were produced by processing the raw MIRI ramp files using \texttt{Dectector1Pipeline}, with the output being run through \texttt{Image2Pipeline} with default parameters. 
Before combining the individual dithers into a final image, we aligned the F770W \texttt{Image2Pipeline} files (*\_cal.fits) to the NIRCam F444W point-source catalog.
We then applied the same astrometric corrections to the remaining MIRI filters. 
Following image alignment, we perform instrumental background subtraction following the method outlined in \citet{bib:Nally2024}.
For each filter, we first median combine all dithers.
Because I~Zw~18 is present in each of these exposures, however, the median-combined backgrounds are not free of structure.
We therefore select unaffected detector rows and columns and map these to the detector plane to produce a model background, which we then subtract from each individual exposure (see \citealp{bib:Nally2024, bib:Dicken2024}).
Finally, the background-subtracted and F444W-aligned individual dither exposures were combined into final images using \texttt{Image3Pipeline} with the \texttt{tweakreg} step turned off.

\subsection{Point-source Extraction} 
\label{sec:pointsources}

The extraction of point sources in I~Zw~18 was accomplished through use of the {\sc starbugii} \citep{bib:Starbug} photometric tool and pipeline.
This software is optimized for JWST observations of crowded stellar fields within complex environments, performing point-source extraction and band merging across multiple observations and wavelengths, utilizing core functions from the python {\sc photutils} \citep{bib:Bradley2022} package.
Examples of programs which have employed {\sc starbugii} for photometric study include \citet{bib:Jones2023}, \citet{bib:Lenkic2024}, \citet{bib:Nally2024}, and \citet{bib:Habel2024}.
The complete set of the relevant {\sc starbugii} parameters and their adopted values used for this program's photometric extractions are listed in Table \ref{tab:starbugparams}.

For each filter in NIRCam, we perform source detection using \texttt{starbug2 $--$detect} on the Stage-3 drizzled image.
We locate sources that are $\sim5\sigma$ above the sky level, and their location is measured by fitting centroids.
Values of the {\em sharpness} and {\em roundness} are assigned en route to calculating a geometry for each point source.
By setting an upper limit on {\em sharpness}, the measured ratio of source peak height to its median pixel value, remaining image artifacts are removed from the catalog, while setting a lower limit assists in removing faint peaks in the dust structure.
Source symmetry is measured by the {\em roundness} parameter.
Most resolved background galaxies, as well as further spurious detections within the dust structure, are removed by limiting the allowed level of asymmetry in the source profiles.
The same routine was used to detect sources in the F770W MIRI image with modified detection parameters.
Because the decrease in image resolution inhibited the detection of robust point-like sources in the longer-wavelength filters (F1000W, F1500W, and F1800W), bright sources in close proximity to other similar objects were instead merged into bright clustered regions that were not analyzed in this study.

Aperture photometry was performed for all four NIRCam filters (F115W, F200W, F356W, and F444W), plus the shortest-wavelength MIRI filter (F770W).
For all NIRCam sources, we use a fixed radius of 1.5 pixels as well as a sky annulus of 3.0 and 4.5 pixels for the inner and outer radii, respectively.
Aperture correction is interpolated between values given in CRDS \texttt{jwst\_nircam\_apcorr\_0004.fits}.
For the MIRI F770W band, the procedure is the same, however the aperture radius is increased to 2 pixels to account for the larger PSFs intrinsic to longer wavelengths, backgrounds are calculated within annuli of 4 and 6 pixels for the inner and outer radii, respectively, and \texttt{jwst\_miri\_apcorr\_0005.fits} is used to calculate aperture corrections.
Through inspection of the data quality array within the aperture of each source, we flag those with saturated, \texttt{DO\_NOT\_USE}, or \texttt{SRC\_JMP} pixels (indicating a jump during detection) with the \texttt{SRC\_BAD} flag in {\sc starbugii}.

We only consider MIRI aperture photometry in the current work as the cruciform structure, which was known to contain up to $\sim$26\% of the flux in the F770W band \citep{bib:Gaspar2021}, was not included in MIRI PSFs simulated by {\sc webbpsf} at the time of the data reduction.

\begin{figure*} 
\centering
\includegraphics[width=0.45\textwidth]{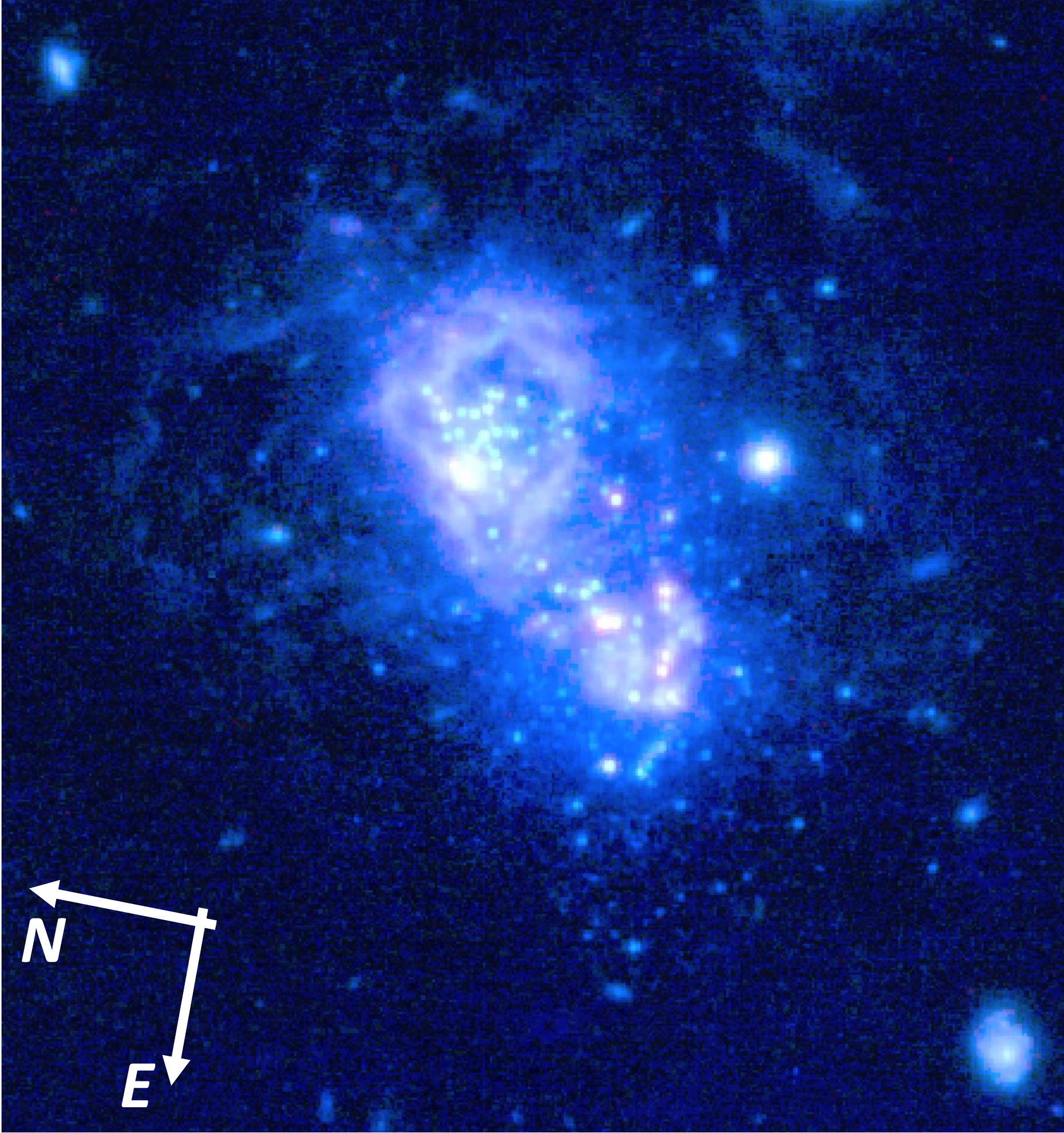}
\includegraphics[width=0.45\textwidth]{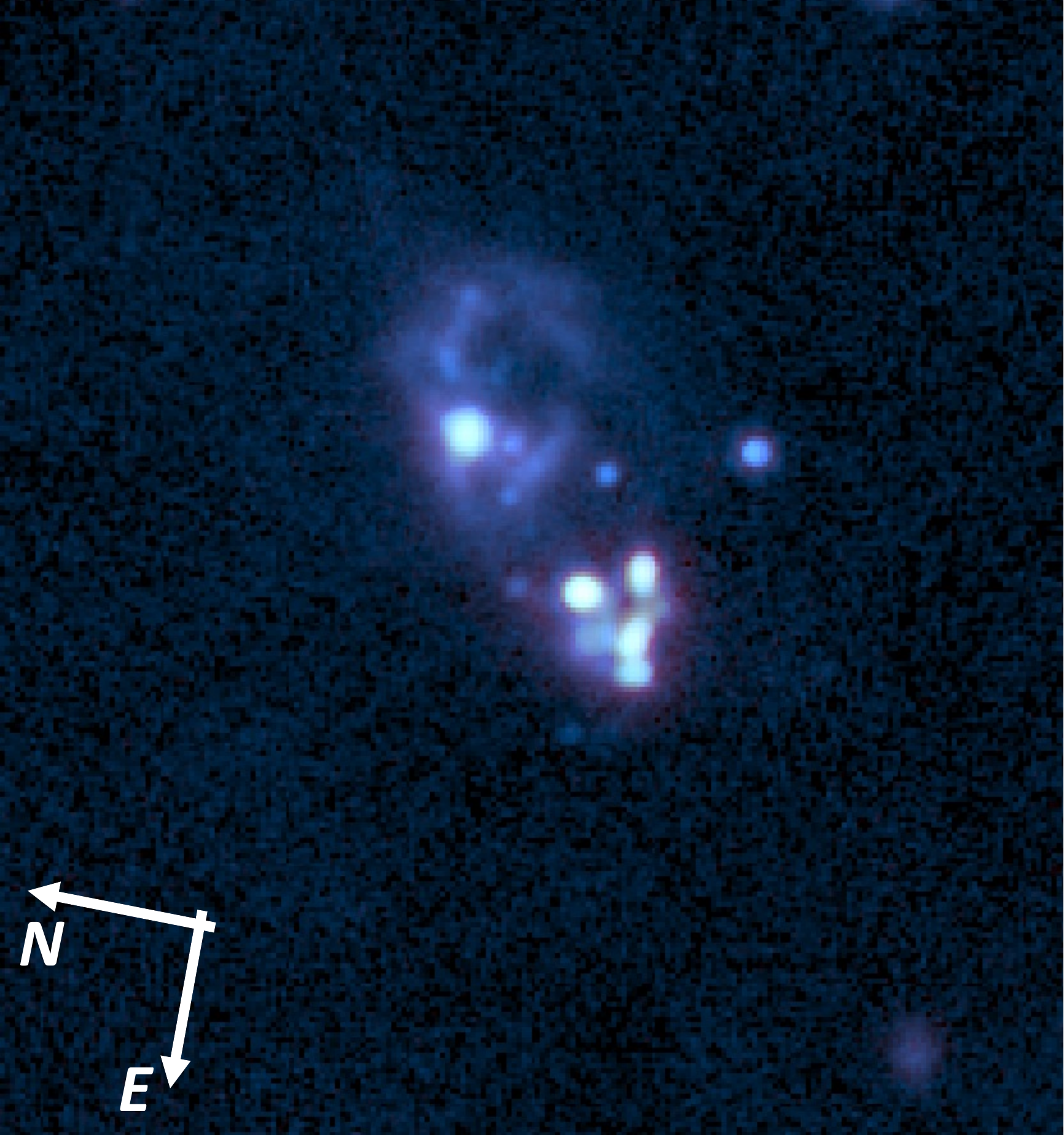}
\caption{
Three-color composite images of I~Zw~18 in near- and mid-IR filters.
A combination utilizing NIRCam F356W, F444W, and MIRI F770W (left) emphasizes the ISM structure of the galaxy in relation to its recent star formation.
The longer-wavelength MIRI F1000W, F1500W, and F1800W filters (right) showcase bulk emission properties of the major SF regions.
Image orientations are such that north is to the left, while east is down, matching that of Figure \ref{fig:prettypicture}.
Located at a distance of 18.2 Mpc, these imaging data present resolutions of 0\arcsec.04--0\arcsec.14 in NIRCam and 0\arcsec.3--0\arcsec.6 in MIRI.
}
\label{fig:prettylong}
\end{figure*}

We perform PSF photometry on the stellar sources of I~Zw~18 on the individual Stage-2 exposures of all four NIRCam filters.
First, the nebulous emission which underlies our imaging data is modeled with the routine \texttt{starbug2 $--$background}.
This places masking apertures of variable size, and fills them with the median pixel value within a local annulus.
Values of the diffuse emission background are then measured for every pixel in the image by averaging all local pixels within a set box size, creating an effective representation of the nebulous emission.
A clean image of the field, with nebulous emission estimate subtracted, is therefore created which enables accurate fitting of PSFs.

The tool {\sc webbpsf} \citep{bib:Perrin2014} version 1.1.1 was then employed to generate a 5 arcsec PSF for each NIRCam detector subarray.
After estimating and subtracting the nebulous background from each single exposure, PSF photometry was performed by running \texttt{starbug2 $--$psf} on the background-subtracted image, at locations defining source positions determined via the source detection.
By leaving the centroid position as a free parameter, both the flux and position of each source are allowed to be fit.
If the position of the new centroid differs from the initial guess determined during the source-detection step by more than 0\arcsec.1, the flux is refit with the source's position held fixed to the initial guess and is flagged with \texttt{SRC\_FIX}, denoting the fit as being of poorer quality.

The calculation and application of instrumental zero-point magnitudes are necessary to calibrate the PSF photometry, as they are not normalized to physical units.
A median difference in source magnitudes measured by the aperture photometry and PSF fitting is calculated using \texttt{starbug2 $--$calc-instr-zp}, allowing for the determination of instrumental zero-point magnitudes, which are then employed to calibrate the PSF photometry from {\sc starbugii} to the AB magnitude system.
These magnitudes are then converted to the Vega system using the reference file \texttt{jwst\_nircam\_abvegaoffset\_0002.asdf}.

\begin{figure*} 
\centering
\includegraphics[width=0.45\textwidth]{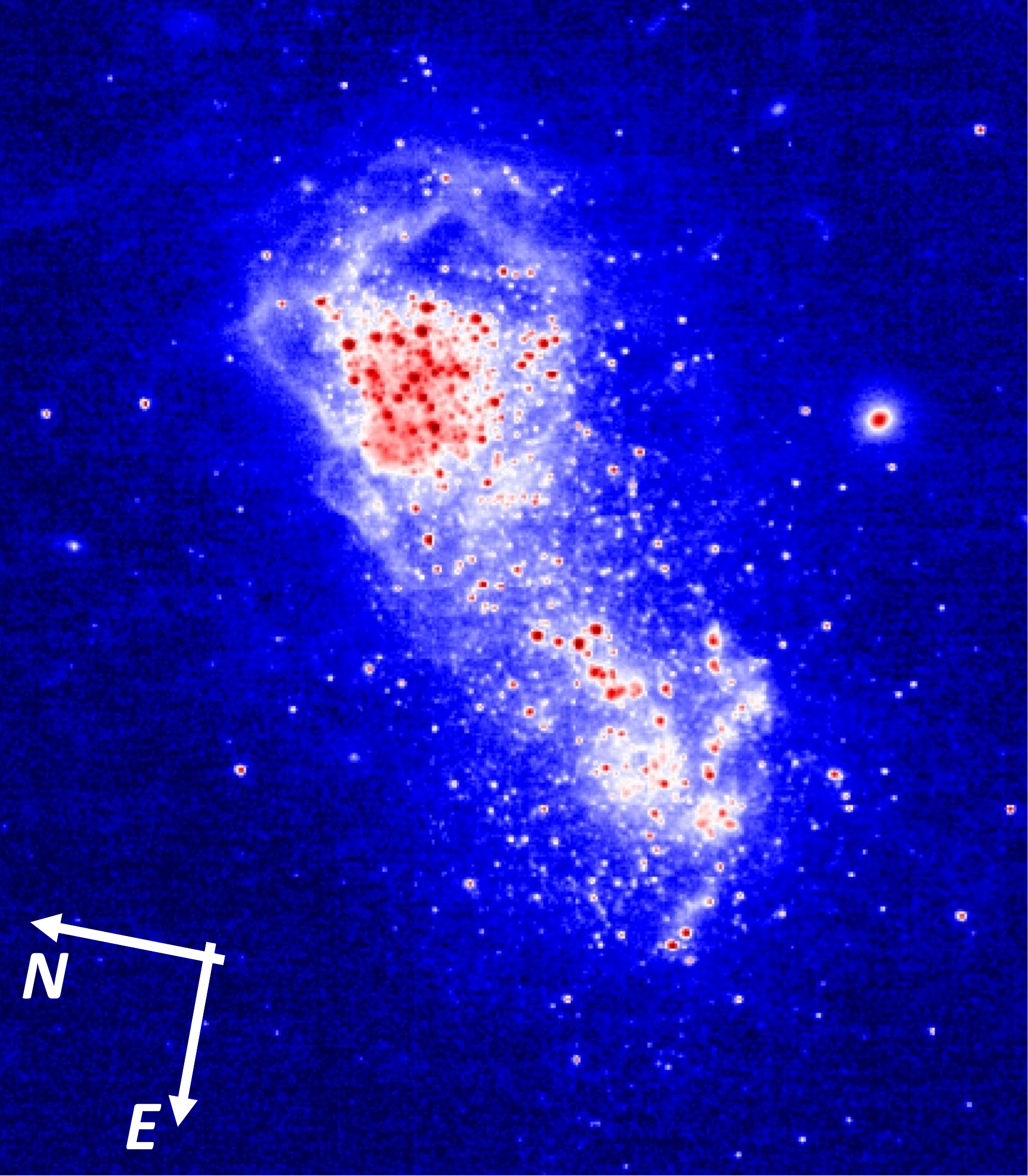}
\includegraphics[width=0.45\textwidth]{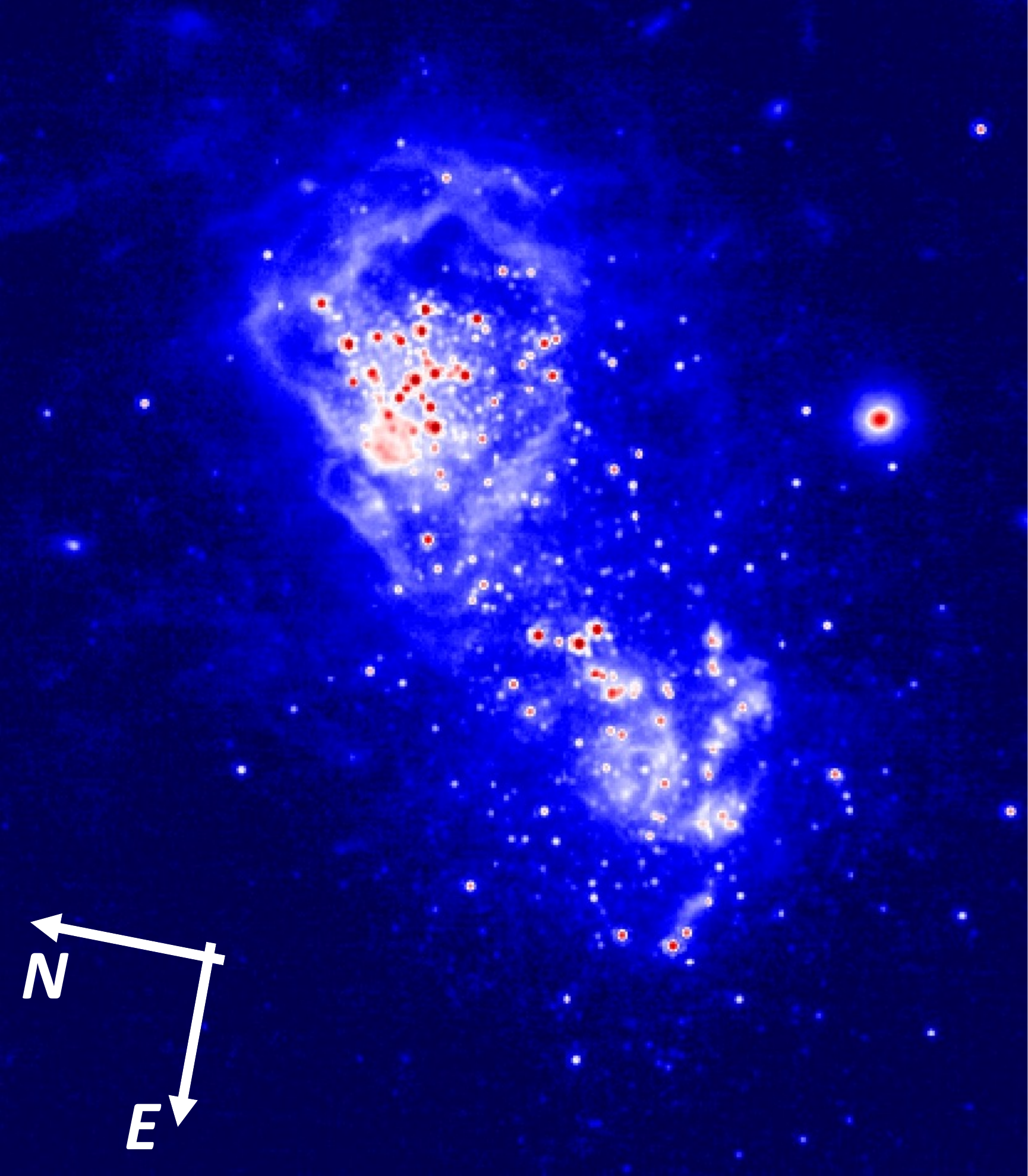}
\includegraphics[width=0.45\textwidth]{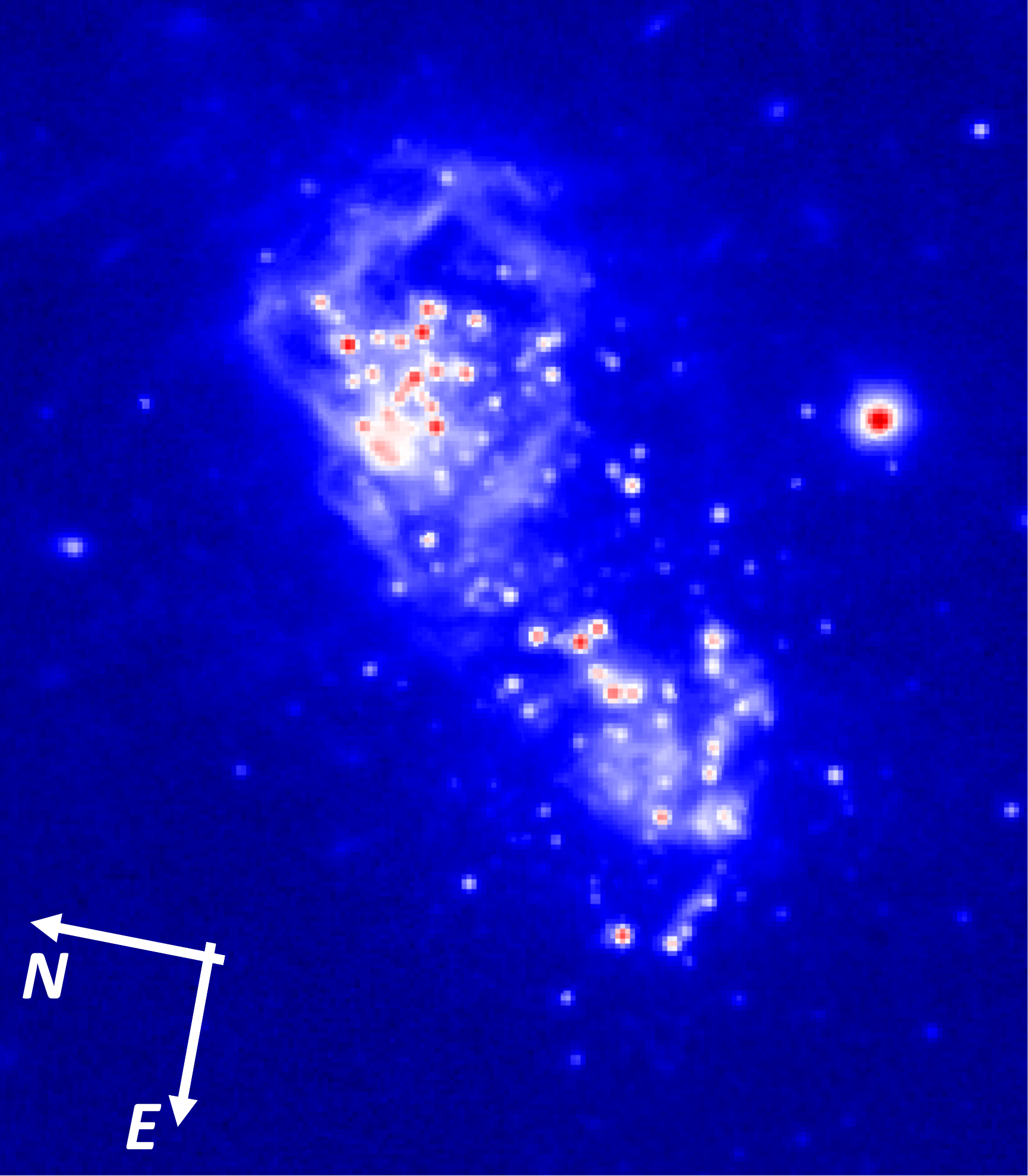}
\includegraphics[width=0.45\textwidth]{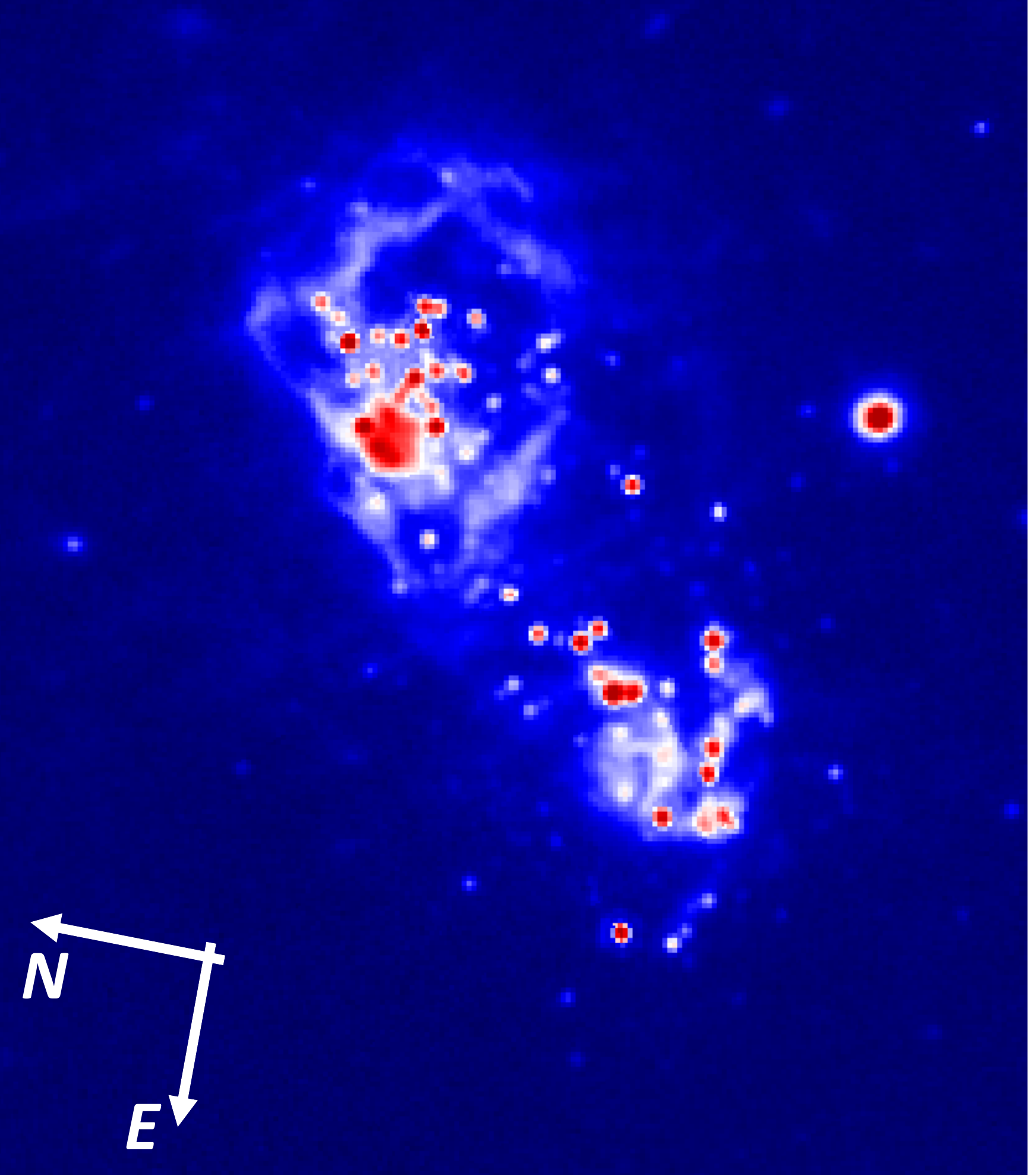}
\caption{
Single-filter false-color images for NIRCam F115W (upper left), F200W (upper right), F356W (lower left), and F444W (lower right).
Clusters of sources emitting in the near-IR reside predominantly in the NW and SE lobes of active star formation.
Detail of the stellar population is revealed from within the obscuring ISM, and IR-bright sources such as AGB stars, RSGs, and bright YSOs become progressively more conspicuous with increasing wavelength.
Image orientations are such that north is to the left, while east is down.
}
\label{fig:NIRCam_images}
\end{figure*}

\begin{figure*} 
\centering
\includegraphics[width=0.45\textwidth]{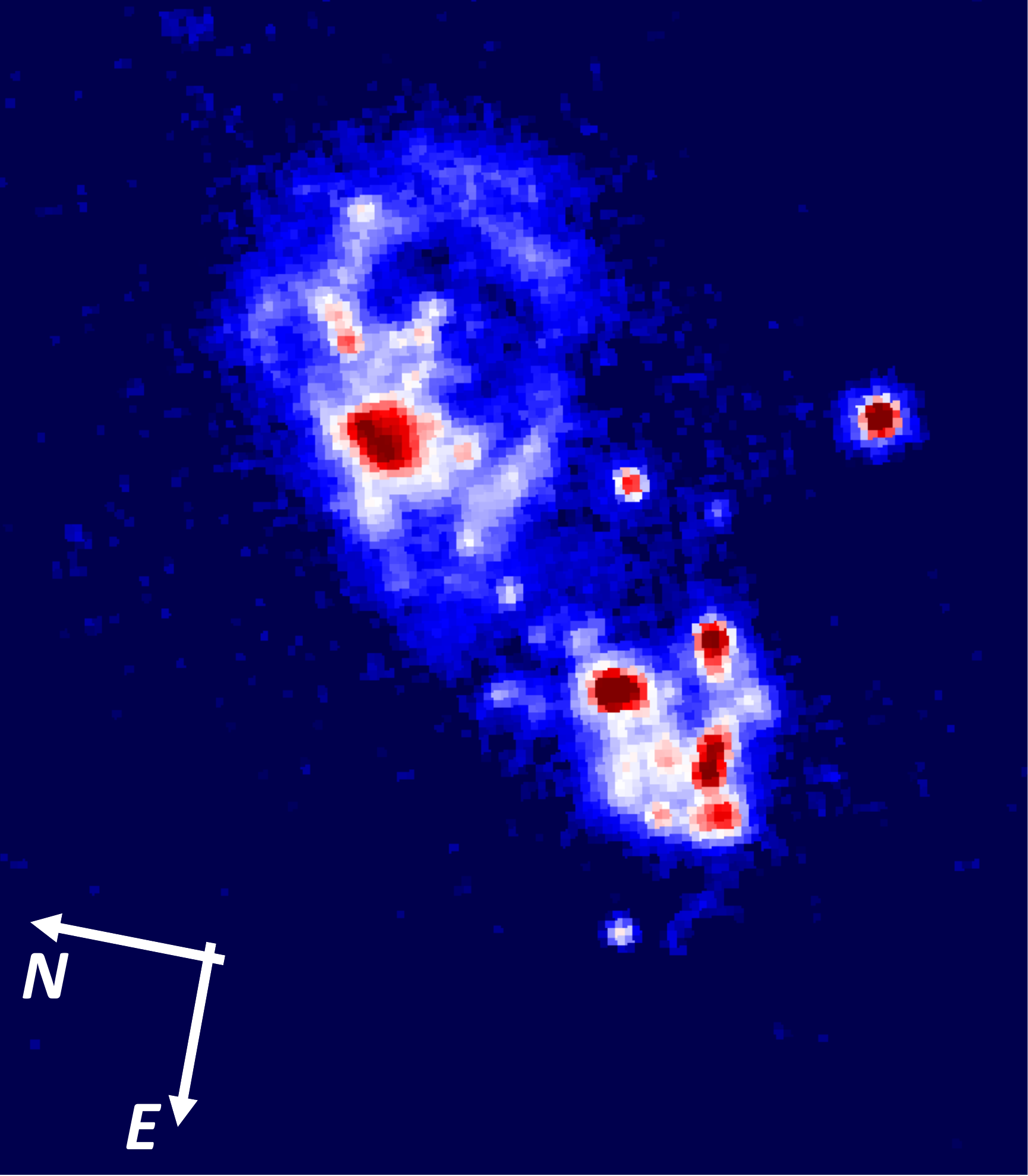}
\includegraphics[width=0.45\textwidth]{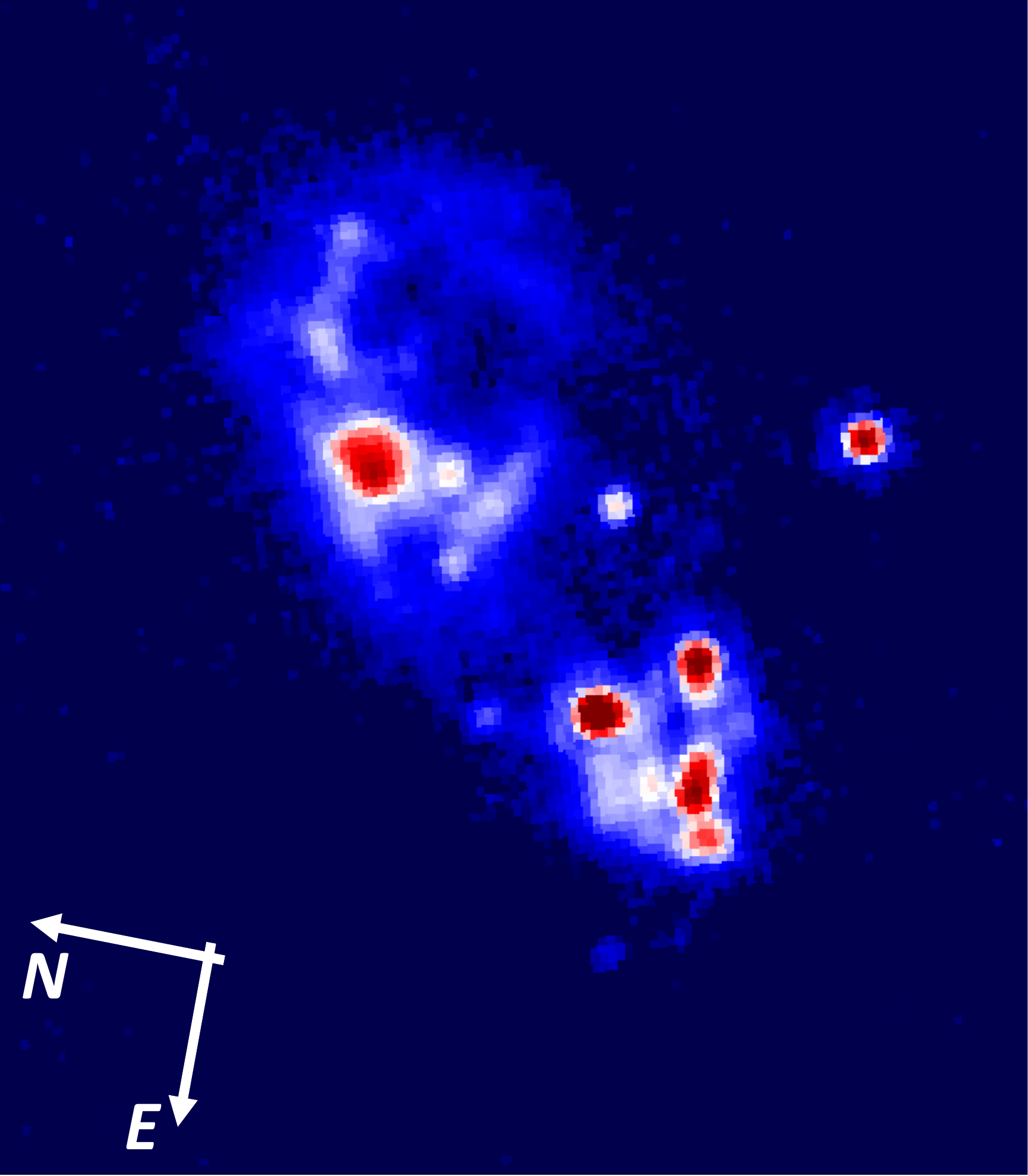}
\includegraphics[width=0.45\textwidth]{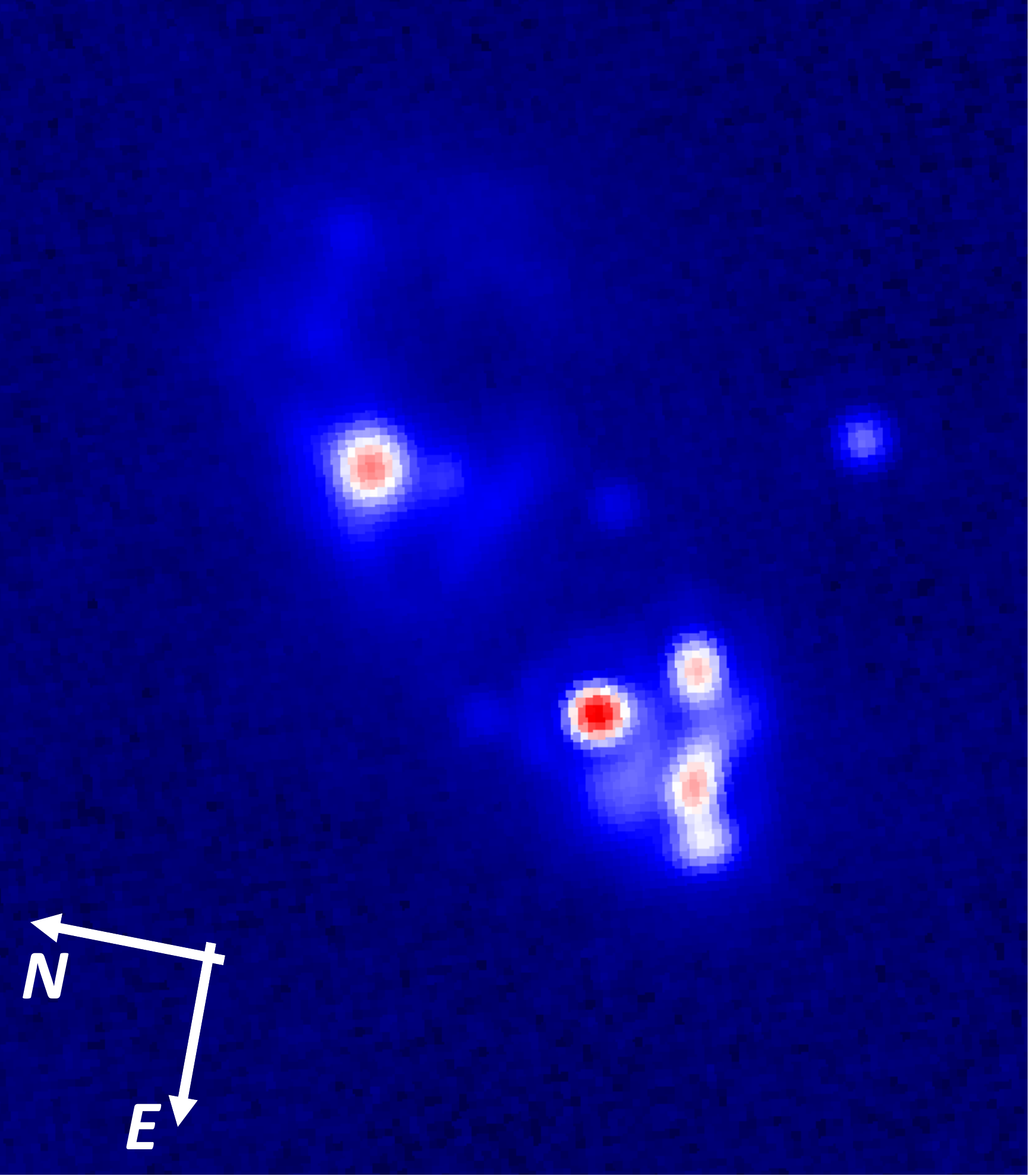}
\includegraphics[width=0.45\textwidth]{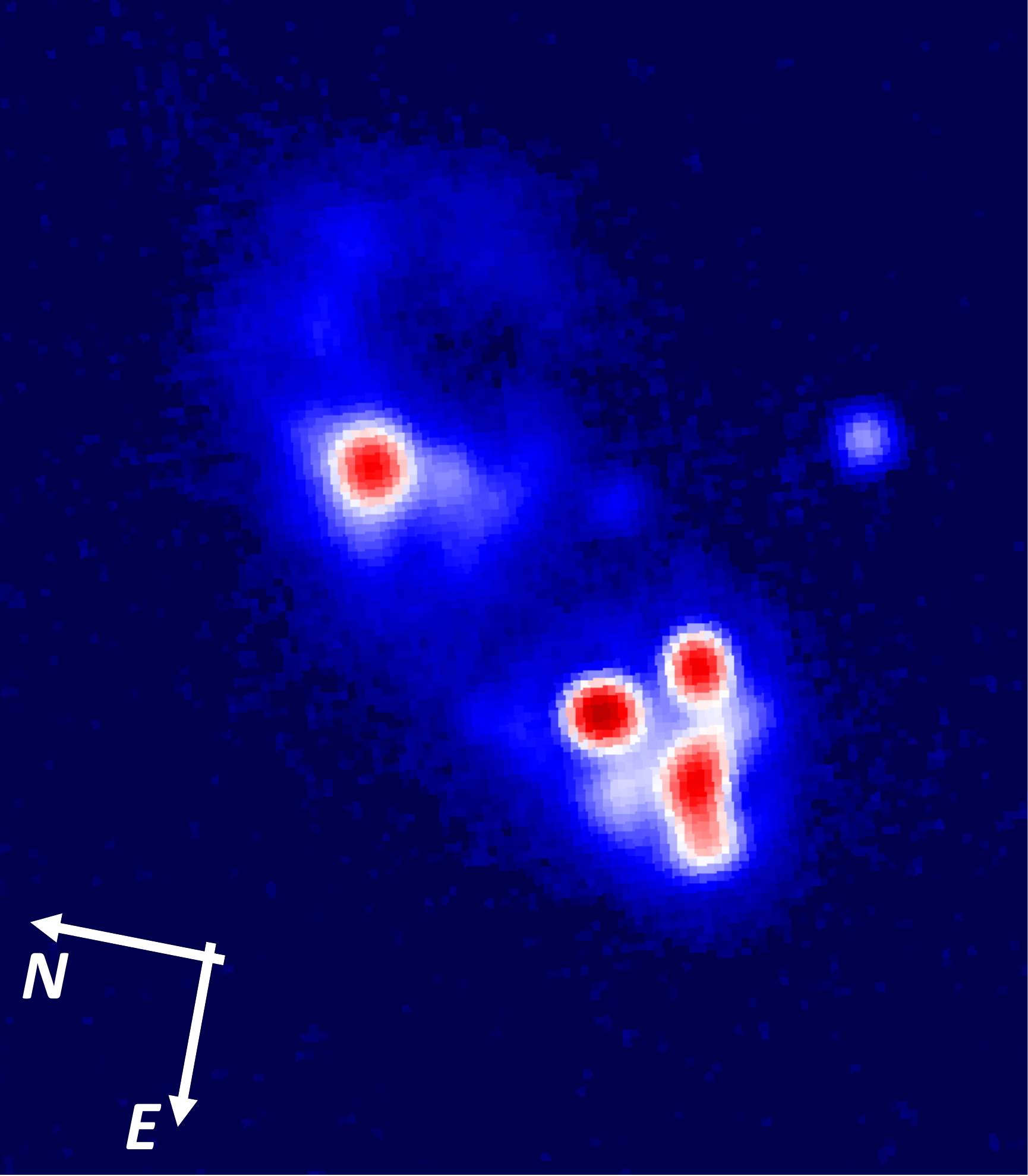}
\caption{
Single-filter false-color images for MIRI F770W (upper left), F1000W (upper right), F1500W (lower left), and F1800W (lower right).
Strong mid-IR emission is correlated with the NW and SE lobes.
Emission from amalgamated IR-bright stellar sources (such as YSO clusters inhabiting massive SF regions) becomes pronounced at longer wavelengths.
Image orientations are such that north is to the left, while east is down.
}
\label{fig:MIRI_images}
\end{figure*}

\subsection{Stellar Source Catalog} 
\label{sec:catalog}

Point-source catalogs for each of the four NIRCam filters were created via the methods described in \S\ref{sec:pointsources}.
A band-matched catalog of point sources is then constructed for I~Zw~18 by merging these catalogs produced for the individual filters using \texttt{starbug2-match $--$band}.
The possibility of mismatching high-quality stellar data with any spurious detector artifacts is mitigated by excluding all sources in the individual filter catalogs with an error on their magnitude greater than 10\%.
While this may limit the potential for completeness of the master catalog, this strategy allows for better reliability overall.

\begin{figure*} 
\centering
\includegraphics[width=1.0\textwidth]{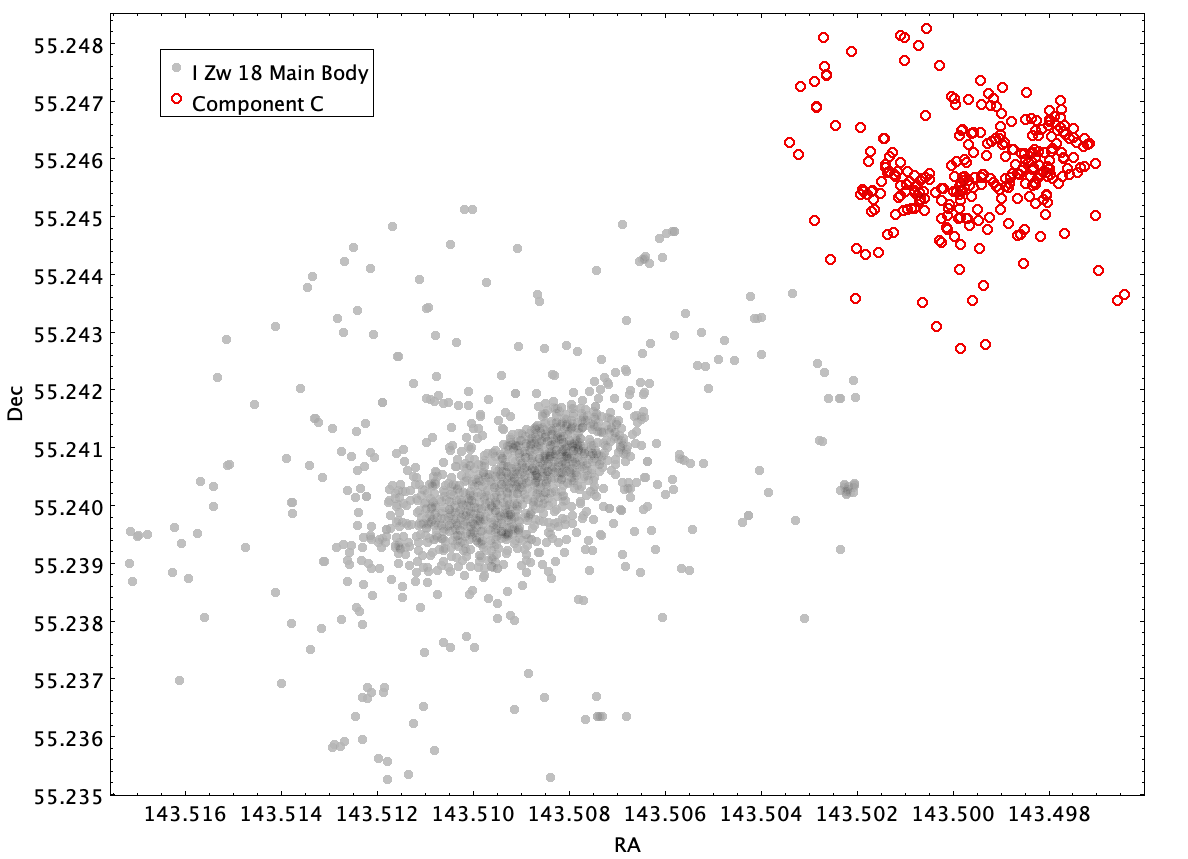}
\caption{
Spatial distribution of NIRCam point sources affiliated with the main body of I~Zw~18 ($n$ = 1,520; gray dots), identified using a 55\arcsec$\times$40\arcsec\ elliptical region centered on the galaxy's main body.
In addition, point sources affiliated with Component C ($n$ = 286; red open circles) are extracted from a 30\arcsec$\times$22\arcsec region $\sim$36\arcsec\ away from the center of I~Zw~18 which avoids artifacts from the detector edges.
}
\label{fig:RADec}
\end{figure*}

First, we consider the catalog made from the shortest-wavelength filter data, for which the smaller PSF FWHM possesses astrometry of the highest certainty.
These sources are then matched using a nearest-neighbor algorithm to sources in the next-shortest-wavelength catalog.
The threshold for separation between filter catalogs increases with increasing wavelength (equivalently, as the size of the PSF increases), as adopting a single separation threshold which is larger than the astrometric uncertainty of, for example, the longest-wavelength NIRCam filter would result in mismatches with, for example, the shortest-wavelength NIRCam filter.
For NIRCam, the SW filters (F115W and F200W) are matched using a threshold of 0\arcsec.06, while the LW filters (F356W and F444W) are matched and added using a threshold of 0\arcsec.10.
Any unmatched sources outside of the separation threshold are then appended to the end of the catalog, before the process is repeated with the next-shortest-wavelength filter data which remains.
We note that, for the nearest-neighbor search method, the position coordinates for the source are adopted from the shortest-wavelength catalog from which said source initially appears.
This allows us to retain the greatest possible positional accuracy for each source.

Due to I~Zw~18's relatively high position with respect to the Galactic plane, along with the galaxies' compact size and small FOV of our observations, the number of contaminating stars inhabiting the foreground is negligible.

\begin{figure*} 
\centering
\includegraphics[width=0.9\textwidth]{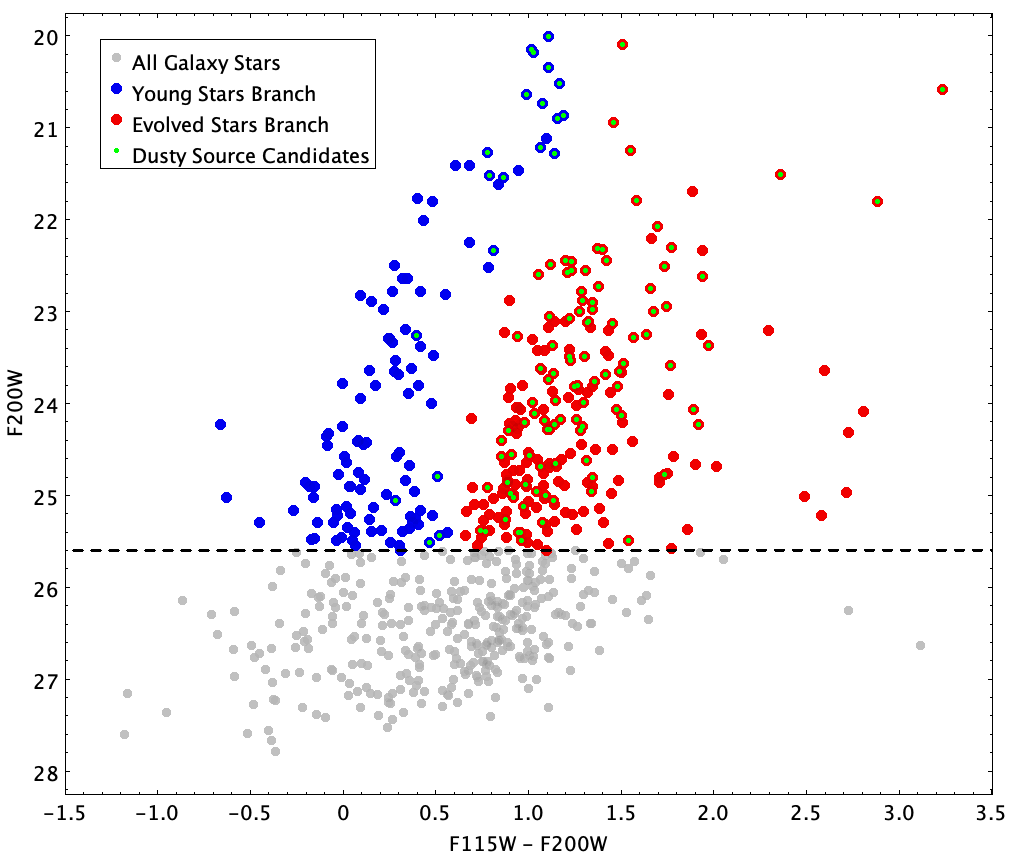}
\caption{
Primary diagnostic CMD for point sources extracted from an elliptical region encompassing the main body I~Zw~18.
This utilizes the two SW NIRCam filters (F200W vs.\ F115W--F200W) and offers the largest potential pool of candidate stars for this study ($n$ = 689).
Stellar source type candidates above the TRGB (F200W $\approx$ 25.6; dashed line) are illustrated as either left-branch (blue) or right-branch (red), representing young or evolved stars, respectively.
Those colored circles over-plotted with a green dot denote characterization as dust-enshrouded candidates (see \S\ref{sec:dusty}).
Dusty objects on the left branch of points are most likely RSGs, while those on the right branch of points are a mixture of AGB stars, RSGs, and bright YSOs.
}
\label{fig:CMD1_populations}
\end{figure*}

To avoid artifacts at the edges of the detectors and unresolved background galaxies impacting our photometric analysis, an elliptical region approximately 55\arcsec$\times$40\arcsec\ aligned along I~Zw~18's semi-major axis was defined.
Only point sources detected within this region were considered to be members of I~Zw~18 in our analysis.
In addition, point sources from Component C were extracted from within a 30\arcsec$\times$22\arcsec\ elliptical region centered on this system, while carefully avoiding the detector edge.
The proximity of Component C to the detector edge makes robust separation between real sources and artifacts challenging; here, we present preliminary results in \S\ref{sec:Component_C}, with a more detailed analysis to follow as the JWST calibrations improve.


\section{Results and Discussion} 
\label{sec:earlyresults}

\subsection{Images} 
\label{sec:images}

Our NIRCam-only image (Figure \ref{fig:prettypicture}) reveals an extreme population of bright, blue, recently formed massive stars, with positions centered upon two primary lobes of star formation, consistent with earlier optical/near-IR studies of I~Zw~18 utilizing HST (e.g., \citealp{bib:HunterThronson1995, bib:Ostlin2000, bib:Cannon2002, bib:IzotovThuan2004}).
An assortment of vivid red stellar sources distributed about the galaxy accentuates the presence of luminous evolved stars.
Also apparent are the large supershells of dust and ionized gas carved out by previous star-formation and SNe activity.
These align with extant maps of emission, including continuum-subtracted \HA\ and \HB\ from \citet{bib:Cannon2002}, which found significant concentrations of dust along such structures.
Resolved background galaxies are apparent throughout the image spanning a variety of orientations and morphology.
Stellar sources associated with the companion system Component C are visible at the top of the image, just left of center.
Gas structure of the ISM is noticeably absent by comparison, consistent with a generally older population of stars bereft of recent star formation (see \S\ref{sec:Component_C} for further analysis).

Images of I~Zw~18 made utilizing MIRI filters (Figure \ref{fig:prettylong}) emphasize the ISM structure of the galaxy via diffuse emission, with loops and filaments tracing a complex gas morphology, and clumpy nature reflecting that of the neutral and molecular gas component (e.g., \citealp{bib:vanZee1998, bib:Cannon2002, bib:Cannon2005a}).
In addition, we see an amalgamation of light originating from the reddest stellar populations, including AGB stars and embedded YSOs, corresponding to the two major SF lobes (the NW and SE components).
These, together with a change in morphology and emission peak with increasing wavelength seen in Spitzer imaging data by \citet{bib:Wu2007}, are resolved here into individual stars, clusters, and ISM structure.

Single-filter false-color images for each of the four NIRCam (Figure \ref{fig:NIRCam_images}) and four MIRI (Figure \ref{fig:MIRI_images}) filters are presented which emphasize details observed in each band.
The spatial distribution of all point sources extracted from both I~Zw~18's main body ($n$ = 1,520; gray dots) and Component C ($n$ = 286; red open circles) are shown in Figure \ref{fig:RADec}.

\subsection{Color--Magnitude Diagrams} 
\label{sec:CMDs}

We construct near-IR CMDs using NIRCam data, which enabled our investigation of the various stellar populations inhabiting I~Zw~18 (Figure \ref{fig:CMD1_populations}).
Our primary diagnostic CMD is F200W vs.\ F115W--F200W, which possesses the highest sensitivity to faint sources and therefore offers the largest potential pool of candidate stars ($n$ = 689).
This combination of filters was employed as a JWST-specific facsimile for standard diagnostic plots, enabling comparison with previous dusty and evolved star studies utilizing Spitzer bands such as SAGE and DUSTiNGS (e.g., \citealp{bib:Blum2006, bib:Boyer2011, bib:Boyer2015a, bib:Boyer2015b}).
In addition, the methods executed by this study mirror recent stellar populations imaging work undertaken with JWST of other extragalactic sources, including NGC 6822 \citep{bib:Lenkic2024, bib:Nally2024} and NGC 346 \citep{bib:Jones2023, bib:Habel2024}.

The structure of points populating this CMD can be roughly described as two primarily vertical columns extending from a common foundation, which terminates as the detection threshold is reached at magnitudes fainter than F200W $\approx$ 28.0.
At this level, the sensitivity limitations of the photometry results in a broad scatter.
Points which make up the blueward (left-side) column of sources represent young stars of the upper main sequence (UMS), with those exhibiting the brightest magnitudes being the most massive.
The redward column of sources is principally populated by evolved stars.
Here, fainter sources are on the red giant branch (RGB), while at brighter luminosities these stars are chiefly He-core-burning RSGs and those on the AGB.
Based on the distance modulus to I~Zw~18, the tip of the red giant branch (TRGB) is estimated to be located at F200W $\approx$ 25.6.
A luminosity function for all galaxy stars (gray) and only those stellar sources populating the evolved star column of sources in this CMD (red) is presented as Figure \ref{fig:lum_func}.
An apparent discontinuity of sources at F200W $\approx$ 25.6 supports this estimation of the TRGB location, though a combination of large photometric uncertainties and small number statistics prevents definitive identification of the TRGB.

\begin{figure} 
\includegraphics[width=\columnwidth]{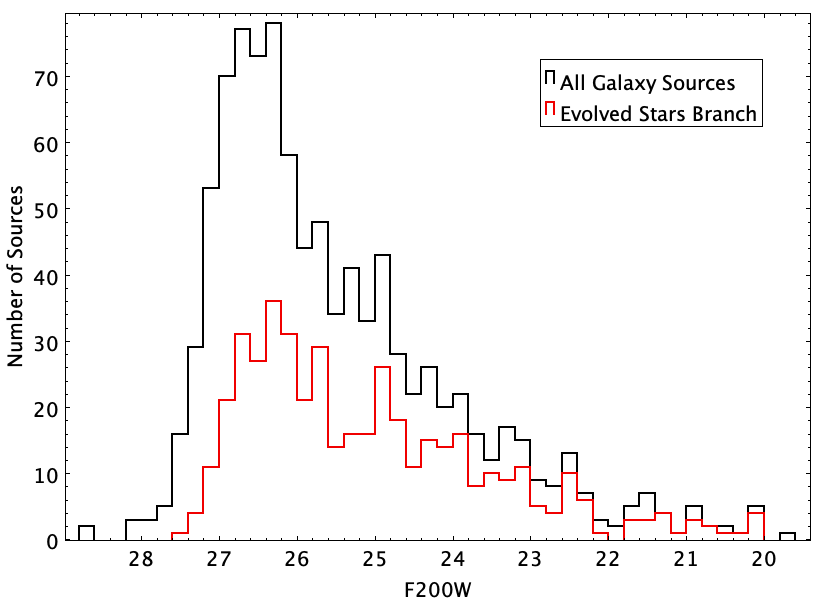}
\caption{
Luminosity function for all galaxy stars (gray) and stellar sources populating the evolved star spur only (red) in the F200W filter, which makes up the \emph{y}-axis of our primary diagnostic CMD.
A discontinuity of detections is apparent at F200W $\approx$ 25.6, supporting the expected location of the TRGB.
}
\label{fig:lum_func}
\end{figure}

In order to assist our preliminary point-source type classifications, we have included for reference \textsc{parsec} isochrones of an XMP stellar population ($Z$ = 0.0003; e.g., \citealp{bib:Bressan2012}) at ages of 1, 10, 25, 50, 100, 250, 500 Myr, 1, 2, 5, and 10 Gyr, overlaid on this CMD (see Figure \ref{fig:CMD1_Isochrones}).
The distribution of sources seen reflects the existence of both older and younger populations; for a thorough examination of the star formation history (SFH) of this galaxy as discerned by the photometric data obtained with NIRCam, see \citet{bib:Bortolini2024}}.
The overall numbers for point-source identifications described in the following subsections from the young star and evolved star branches, including their respective dust-enshrouded candidates (\S\ref{sec:dusty}), plus YSO and YSO cluster candidates (\S\ref{sec:youngstellarobjects}), are summarized in Table \ref{tab:pointsourcedetections}.

\begin{figure*} 
\centering
\includegraphics[width=0.9\textwidth]{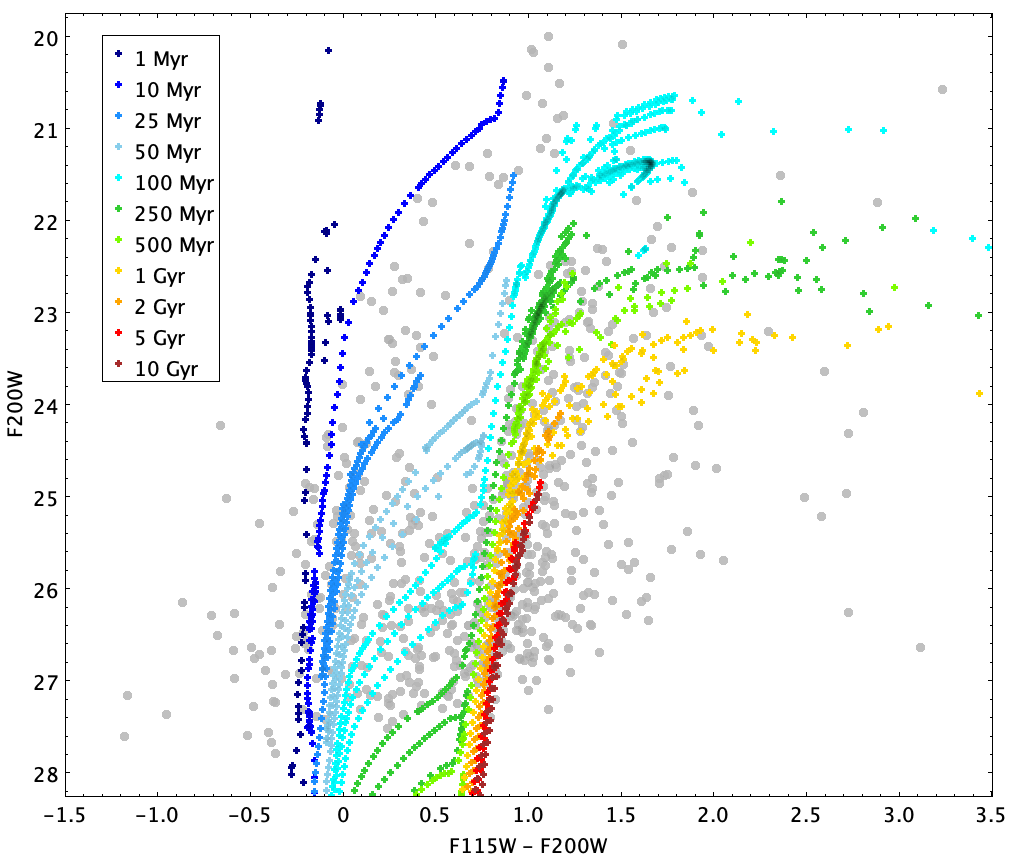}
\caption{
F200W vs.\ F115W--F200W CMD with overlaid \textsc{parsec} isochrones of an XMP stellar population ($Z$ = 0.0003) at ages of 1 Myr, 10 Myr, 25 Myr, 50 Myr, 100 Myr, 250 Myr, 500 Myr, 1 Gyr, 2 Gyr, 5 Gyr, and 10 Gyr.
}
\label{fig:CMD1_Isochrones}
\end{figure*}

\subsubsection{Dusty Evolved Star Candidates} 
\label{sec:dusty}

Identification of dust-enshrouded stellar source candidates in I~Zw~18 using IR photometry was accomplished by examining a CMD employing our most widely separated NIRCam filter data (F444W vs.\ F115W--F444W; Figure \ref{fig:dusty_CMDs}).
At these longer wavelengths, the numbers of detected point sources ($n$ = 183) diminish in comparison to our primary diagnostic CMD.
This large color baseline, however, affords a robust separation between dusty and nondusty stellar types:
Any source detection with a color value F115W--F444W $>$ 0.0 may be considered to be at least somewhat enshrouded in dust.
In order to account for photometric uncertainties, multiple ages, variance in metallicities, and other broadening effects, we have elected to implement a more conservative color cut at F115W--F444W = 0.5.
Our dust-enshrouded source candidates therefore require detection in all three of the F115W, F200W, and F444W filters; a star becomes ineligible for inclusion in our high-confidence catalog if it is missing from any one filter's detected source list.
While the overall effect is that we are possibly missing some real dust-enshrouded source candidates in our study, we are opting to remain conservative.
We also note that, while the MIRI F770W filter provides for a yet-longer color baseline, the paucity of sources detected via the F115W--F770W color for our data diminishes its utility as a diagnostic tool.
At the distance to I~Zw~18 (18.2 Mpc), the resolution and sensitivity of MIRI F770W are substantially less than that of NIRCam F444W, such that including this baseline could result in blending in the F770W filter, leading to source mismatches between catalogs.

The F444W vs.\ F115W--F444W CMD (Figure \ref{fig:dusty_CMDs}) employed to characterize dust-enshrouded source candidates in I~Zw~18 bears resemblance to the [4.5] vs.\ $K$--[4.5] CMD used by the \citet{bib:Dell'Agli2019a} study of Sextans A, a nearby ($D$ = $\sim$1.32 Mpc; \citealp{bib:Dolphin2003}) low-metallicity (\abun\ $\approx$ 7.6; \citealp{bib:Skillman1989c, bib:Pilyugin2001}) SF dwarf galaxy (see their Figures 2 and 3).
While \citet{bib:Dell'Agli2019a} utilized this plot alongside evolutionary model tracks of metal-poor stars in order to make distinctions between evolved star populations (RSGs, O-rich AGBs, C-rich AGBs), we do not attempt to make equivalent early distinctions.
Because of the larger distance to I~Zw~18, our data suffers from higher photometric uncertainties as a consequence of crowding effects.

\begin{figure} 
\centering
\includegraphics[width=\columnwidth]{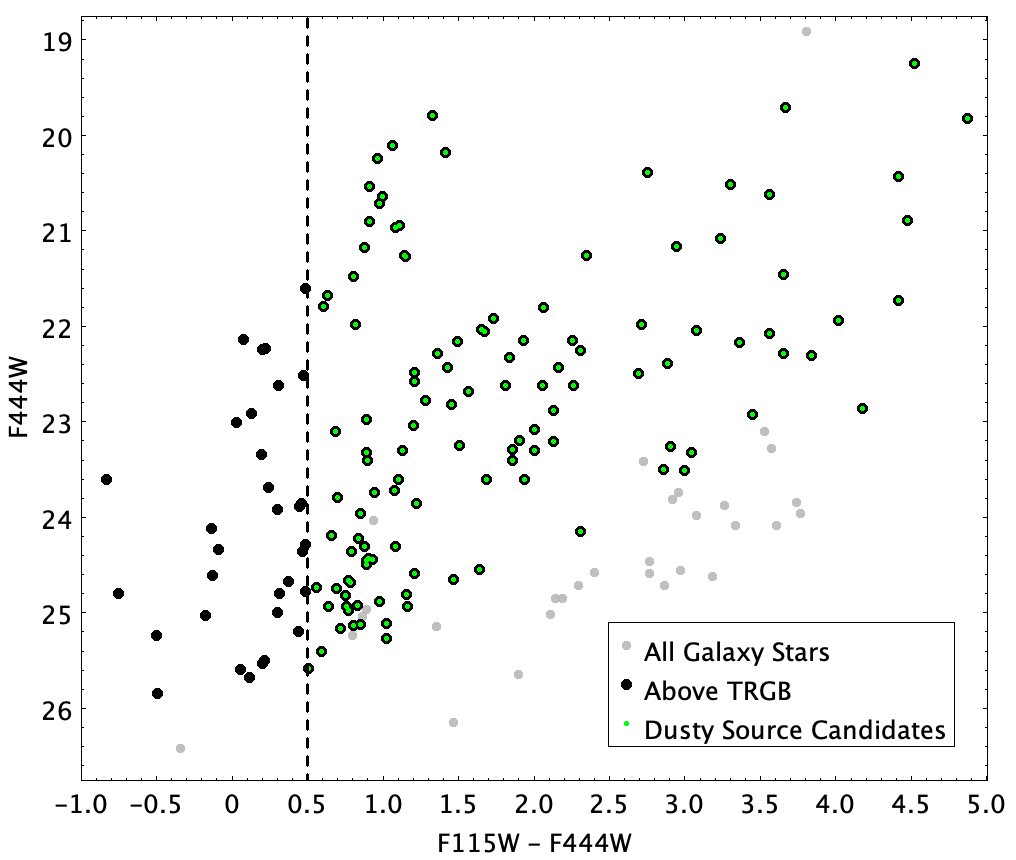}
\caption{
CMD constructed with our widest-possible color baseline (F444W vs.\ F115W--F444W) NIRCam data, used to characterize dust-enshrouded object candidates.
Sources which were found above the TRGB in our primary diagnostic CMD (F200W vs.\ F115W--F200W; see Figure \ref{fig:CMD1_populations}) are denoted as black circles, while those below are shown in gray.
Dusty source candidates in I~Zw~18 are those found both above the TRGB and possessing a color value F115W--F444W $>$ 0.5 (dashed line), and are presented here as black circles over-plotted with green dots ($n$ = 119).
}
\label{fig:dusty_CMDs}
\end{figure}

For our list of dust-enshrouded source candidates identified from this CMD, we additionally required each detection to have been found in the F200W vs.\ F115W--F200W CMD located above the TRGB (F200W = 25.6).
We note that, while some additional bona fide dust-enshrouded source candidates are likely to lie below this TRGB, we have elected to exclude them here for the sake of remaining conservative with our estimates, and to reflect larger photometric uncertainties at these magnitudes.
For objects inhabiting the right-side (evolved stars) branch of the F200W vs.\ F115W--F200W CMD, those which are additionally identified as being dusty in the F444W vs.\ F115W--F444W CMD are expected to be AGB stars, RSGs, and bright YSO clusters.
The fewer dusty sources of the left-side (young stars) branch of the F200W vs.\ F115W--F200W CMD are likely RSGs.
In total, we find $n$ = 119 dusty source candidates in the F444W vs.\ F115W--F444W CMD, which are signified both in Figures \ref{fig:CMD1_populations} and \ref{fig:dusty_CMDs} as overlaid green dots.
Of these, $n$ = 99 belong to the evolved stars branch, while $n$ = 20 belong to the young stars branch.
Overall, of the $n$ = 226 sources populating the evolved stars branch, the $n$ = 99 robust dust-enshrouded source candidates represent $\sim$43.8\% of the total.
While the $n$ = 20 dust-enshrouded source candidates on the young stars branch constitute only $\sim$17.2\% of the total ($n$ = 116), these include the majority of bright sources expected to be RSGs.

\begin{figure*} 
\centering
\includegraphics[width=0.75\textwidth]{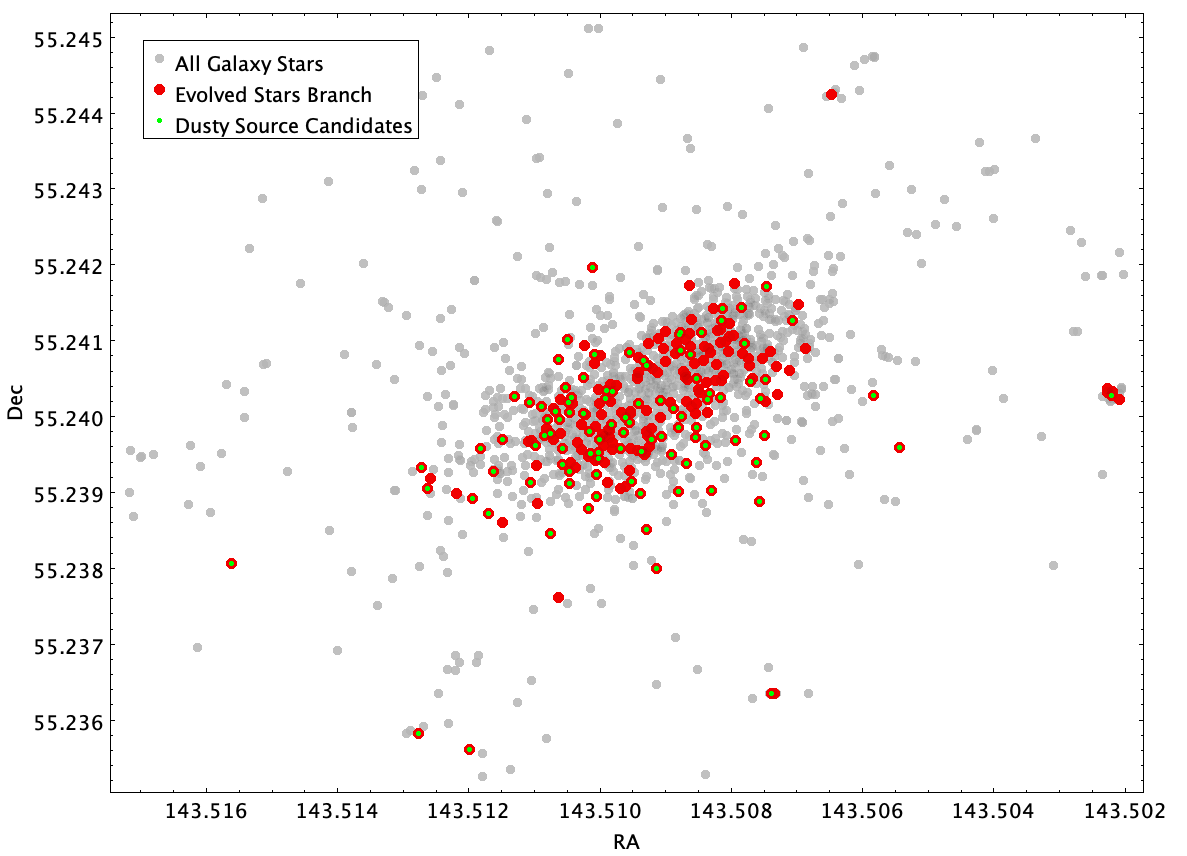}
\includegraphics[width=0.75\textwidth]{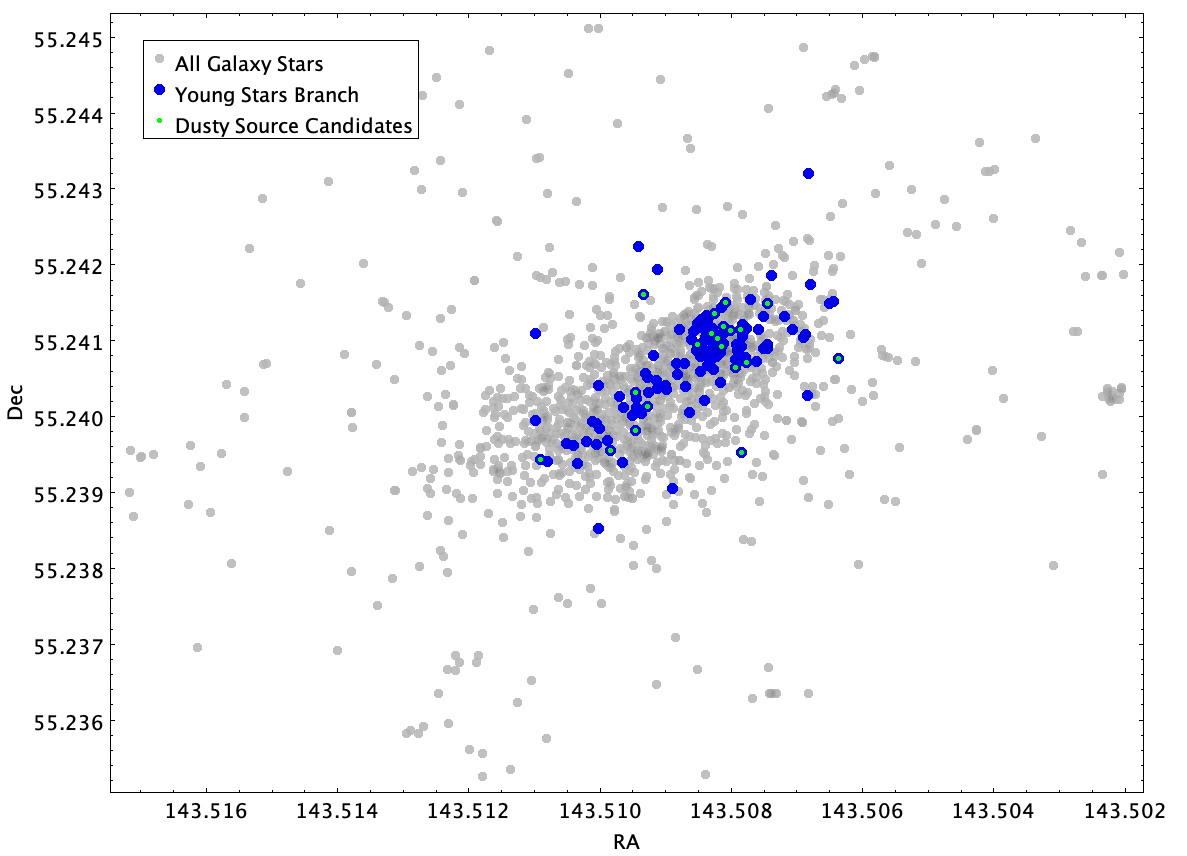}
\caption{
Spatial distribution plots for stellar sources populating the right branch ($n$ = 226; \emph{top}) and left branch ($n$ = 116; \emph{bottom}) of the F200W vs.\ F115W--F200W CMD, as found via photometric analyses of I~Zw~18.
Right-branch points indicated as red circles are a mix of evolved source candidates, including both O- and C-rich AGB stars as well as RSGs, in addition to bright YSOs.
Left-branch points indicated as blue circles are predominantly hot young stars, including those populating the UMS, and bright RSGs.
The dustiest of these objects, identified in the F444W vs.\ F115W--F444W CMD, are marked as over-plotted green dots in both cases ($n$ = 99 and $n$ = 20, respectively).
These sources are found chiefly clustered around the galaxy's two major lobes of star formation (NW and SE).
}
\label{fig:spatdist}
\end{figure*}

Stars inhabiting the AGB represent the final evolutionary stage of low- to intermediate-mass ($\sim$0.6--10 $M_{\odot}$) MS progenitors, and are believed to be a major contributor of interstellar dust.
Of interest to this observing program in particular are thermally pulsing (TP-AGB) stars which lie above the TRGB.
We may classify these AGB stars into two major categories:\
oxygen- and carbon-rich (O-rich and C-rich, respectively).
MS progenitors with masses $\sim$3--8 $M_{\odot}$ evolve to become AGB stars which are rich in oxygen, but have short lifespans ($t_{\mathrm{life}}$ $\sim$ 30--200 Myr).
While they are capable of producing dust relatively quickly after having formed, they are not expected to enrich the ISM with \emph{s}-process elements in metal-poor environments to the same degree as AGB stars with lower initial masses ($\sim$1--3 $M_{\odot}$), which are more prevalent (e.g., \citealp{bib:Lugaro2003, bib:KarakasLugaro2016, bib:Karakas2010}).
Fewer \emph{s}-process elements may result in fewer nucleation sites for dust growth in the ISM.
These stars possess longer lifespans ($t_{\mathrm{life}}$ $\sim$200 Myr--3.6 Gyr; \citealp{bib:KarakasLattanzio2003, bib:Ventura2013, bib:Boyer2015a}), during which they experience repeated third dredge-up episodes \citep{bib:Kwok2000} that lead to their becoming increasingly carbon-enriched \citep{bib:KarakasLattanzio2014}.
O- and C-rich TP-AGB stars may therefore be distinguished from one another based on the dominant chemistries of their photospheres.
If the value of the ratio of carbon to oxygen (C/O) is $>$1, the AGB star is C-rich, while a C/O value $<$1 is O-rich.
Dust species born from these populations depend upon the dominant element, such that carbonaceous dust grains are produced by C-rich AGB stars, while silicates are produced by O-rich AGB stars.

The stellar mass value at which the distinction between O- and C-rich AGB stars occurs exhibits a dependence on metal abundance, decreasing from $\sim$3 $M_{\odot}$ as metallicity gets lower:
The condition of C/O $>$ 1 is reached more easily for stars in environments with lower abundances of oxygen.
In the case of extremely low metallicities, such as that exhibited by I~Zw~18, the threshold for producing more C-rich stars may occur as low as $\sim$2.5 $M_{\odot}$ \citep{bib:Dell'Agli2018}.

Without the wider color baselines afforded by MIRI photometry, robust segregation between O- and C-rich AGB stars is not possible.
Many of the evolved star branch sources identified in the F200W vs.\ F115W--F200W CMD (Figure \ref{fig:CMD1_populations}; red points), however, are expected to be TP-AGB stars.
In addition, considering the extremely low metallicity of I~Zw~18, many of these TP-AGB candidate stars are expected to be C-rich.
The distribution of dust-enshrouded evolved stars (red points with overplotted green dots) permeates the parameter space occupied by the F200W vs.\ F115W--F200W CMD's evolved star branch, suggesting the presence of many dusty C-rich AGB star candidates in I~Zw~18.

The locations of the evolved star branch sources in this galaxy are illustrated in Figure \ref{fig:spatdist} (top) as red points, with dusty source candidates signified with overplotted green dots.
Many of these objects are anticipated to be TP-AGB stars (other possibilities include massive YSOs, however positive classification will require spectral energy distribution model fitting; see \S\ref{sec:youngstellarobjects} for more details), with the dustiest among them appearing to cluster around the large SF region located in the lower left (the SE lobe).
Furthermore, these AGB star candidates show consistency with identifications from past studies utilizing HST (e.g., \citealp{bib:OstlinMouhcine2005}).

Stars identified as RSGs are a late evolutionary phase of higher-mass ($\gtrsim$10 $M_{\odot}$) MS progenitors which trace regions of recent star formation.
These short-lived, cool, very luminous objects occupy color space coincident with TP-AGB stars, and are expected to be found alongside them in the F200W vs.\ F115W--F200W CMD's evolved star branch.
The majority of the brightest sources populating the left-side branch (Figure \ref{fig:CMD1_populations}; blue points), however, are denoted as being candidate dust-enshrouded stars.
These are therefore likely RSG candidates, as well, where they are illustrated in Figure \ref{fig:spatdist} (bottom) as blue points with overplotted green dots.
We find that these dusty left-branch RSG candidates are strongly clustered around the large SF region located in the upper right (the NW lobe), opposite to that of the dusty TP-AGB stars discussed earlier.
This broad demarcation of differently aged sources suggests that the bulk of the most recent star-formation episode of the SE lobe, populated predominantly with slower-evolving C-rich AGB stars, took place longer ago than that of the NW lobe, which is still inhabited by younger sources such as RSGs.

\subsubsection{Young Stellar Object Candidates} 
\label{sec:youngstellarobjects}

The active star formation taking place in I~Zw~18 makes it an excellent local analog for conditions which pervaded the early Universe, when heavy-element enrichment levels had yet to reach what is found today.
YSOs are birthed in active star formation regions and exhibit strong IR excess as light is absorbed and reemitted by cool, dusty envelopes and accretion disks \citep{bib:Meixner2013, bib:Seale2014}.
Identification of YSOs in I~Zw~18 represents an improvement in our understanding of star formation mechanisms at the very low levels of chemical enrichment reminiscent of the early Universe, as resolved studies in higher-redshift galaxies that are comparably metal-poor are observationally precluded in all but a few rare circumstances.
These YSOs trace regions of active star formation and emit even more strongly at long wavelengths than dusty AGB stars, particularly at the earliest stages of their lifetimes.
This disk and its circumstellar envelope disperses as the protostar evolves \citep{bib:Robitaille2006}, and so by consequence its dominant emission shifts from mid- to near-IR.

We identify YSO candidates in I~Zw~18 as point sources exhibiting very red colors, similar to those of dust-enshrouded TP-AGB stars.
This is accomplished by implementing the strategy featured in \citet{bib:Jones2023} and \citet{bib:Habel2024} for JWST study of YSOs in the SMC SF region NGC 346.
The selection criteria require the source to be redward of the evolved stars branch from the F200W vs.\ F115W--F200W CMD (with color F115W--F200W $>$ 1.2).
Additionally, the YSO candidates must satisfy the condition of F200W--F444W $>$ 1.0, which was used for NGC 346 to establish the presence of IR excess.
Finally, in order to correct for source contamination from background galaxies, which possess similar IR colors, we perform a final visual inspection of the YSO candidates in the NIRCam F115W image to reject extended objects.
We present these results in the color-color diagram (CCD) as Figure \ref{fig:CCD} (left), with $n$ = 15 point sources plotted in green which inhabit the dashed line box illustrating these parameters.
The location of these YSO candidates on the F200W vs.\ F115W--F200W primary diagnostic CMD (Figure \ref{fig:CCD}; right) emphasizes the characteristic very red colors, as well as some magnitudes fainter than the TRGB (F200W = 25.6).
We note that due to limitations of resolution and sensitivity at the distance to I~Zw~18, any YSO candidates putatively identified in this study are likely to be unresolved clusters of several young stars.

\begin{figure*} 
\centering
\includegraphics[width=0.49\textwidth]{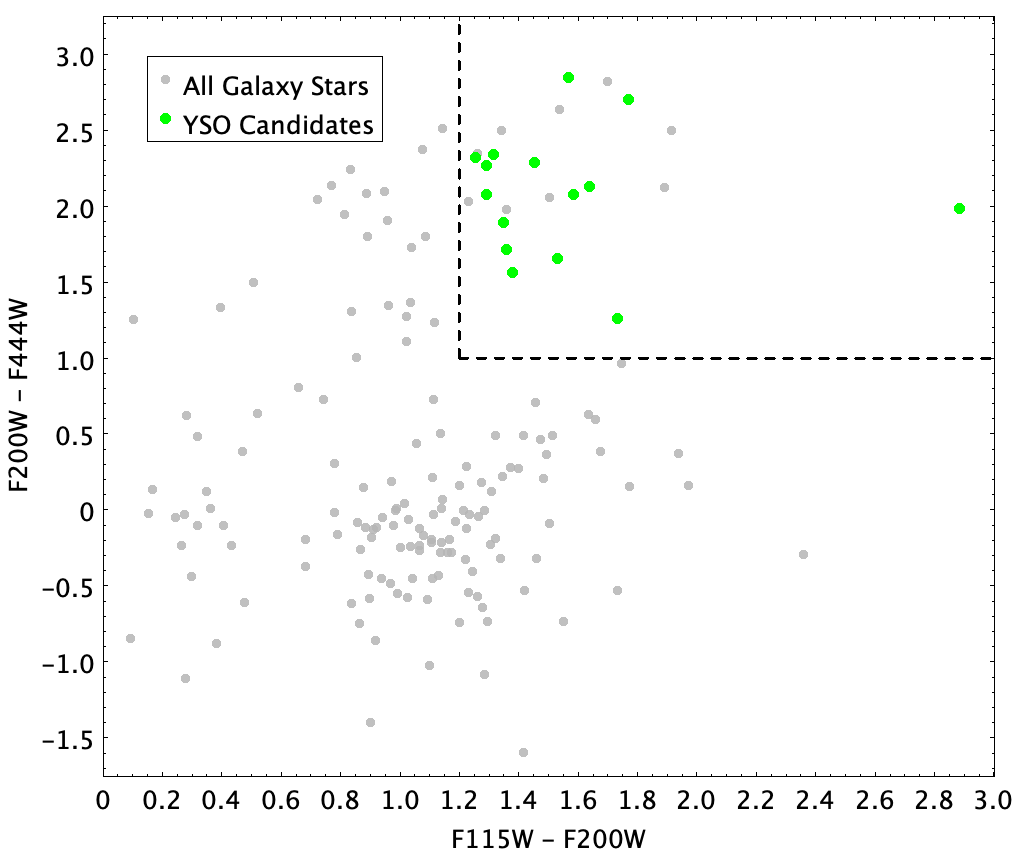}
\includegraphics[width=0.49\textwidth]{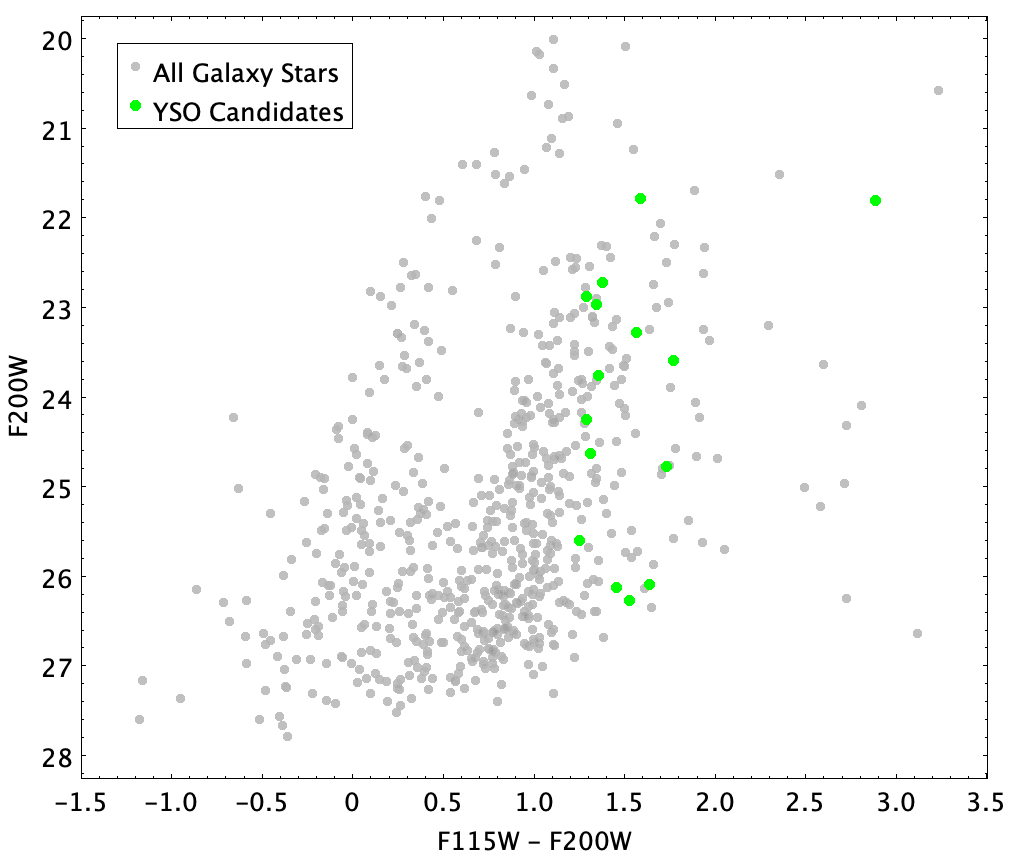}
\caption{
Unresolved YSO candidates ($n$ = 15) identified in I~Zw~18 presented as green circles.
The F200W--F444W vs.\ F115W--F200W CCD (left) illustrates the color-selection criteria implemented by the JWST studies of \citet{bib:Jones2023} and \citet{bib:Habel2024} of the SF region NGC 346 in the SMC.
These same sources are over-plotted on the F200W vs.\ F115W--F200W primary diagnostic CMD, demonstrating positioning redward of the evolved stars branch, and to fainter magnitudes than the TRGB.
}
\label{fig:CCD}
\end{figure*}

The spatial distribution of YSO candidates in I~Zw~18 is illustrated as Figure \ref{fig:spatdist_YSO}, shown again as green circles.
Several sources appear to be grouped together within the SE lobe, consistent with a recent bout of star formation.
YSO candidates are also situated along the outskirts of the NW lobe, where peak star formation activity is expected to have taken place longer ago.
The locations of these sources appear to correlate with the strongest emission seen in MIRI F770W imaging (see Figure \ref{fig:MIRI_images}).

\begin{figure*} 
\centering
\includegraphics[width=0.75\textwidth]{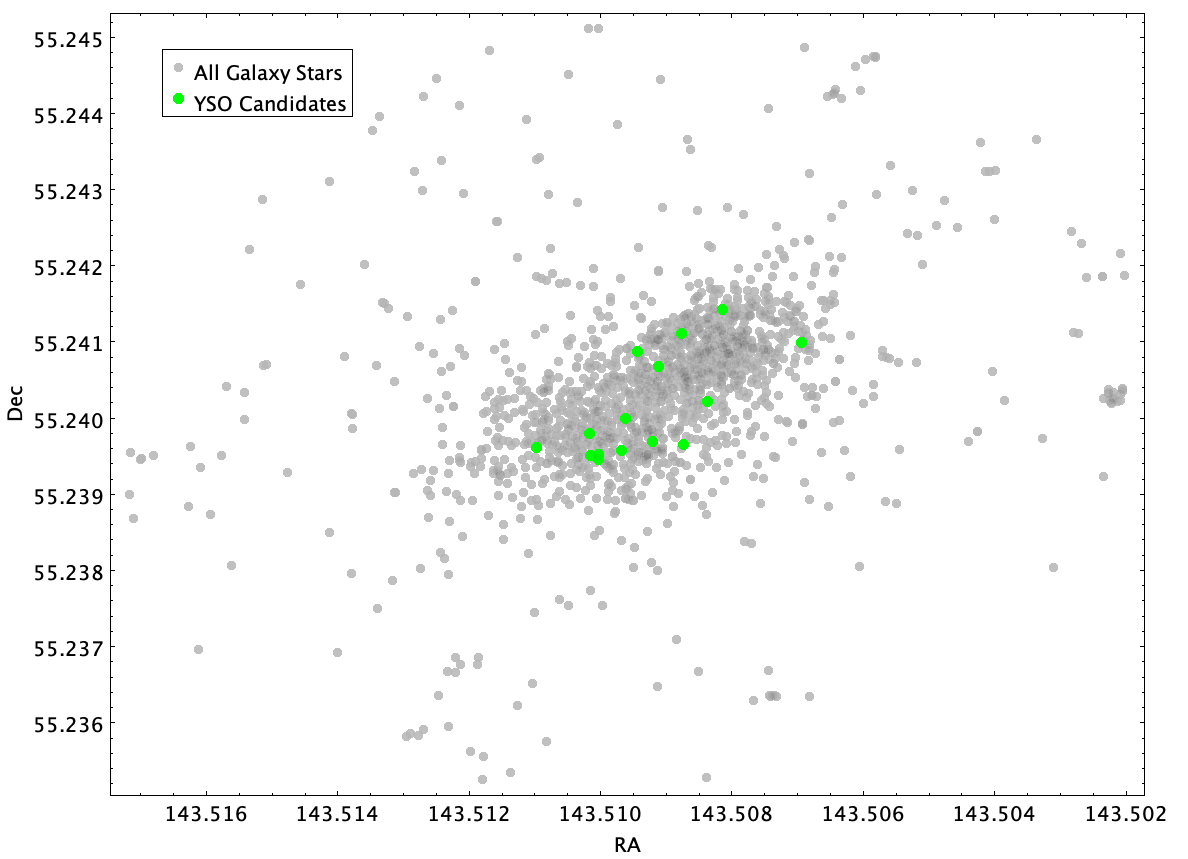}
\caption{
Spatial distribution plot for YSO candidates (green circles) in I~Zw~18 identified via analysis of the F200W--F444W vs.\ F115W--F200W CCD.
We find several sources grouped in the SE lobe of recent star formation, and along the outskirts of the older NW lobe.
}
\label{fig:spatdist_YSO}
\end{figure*}

\begin{figure*} 
\centering
\includegraphics[width=0.9\textwidth]{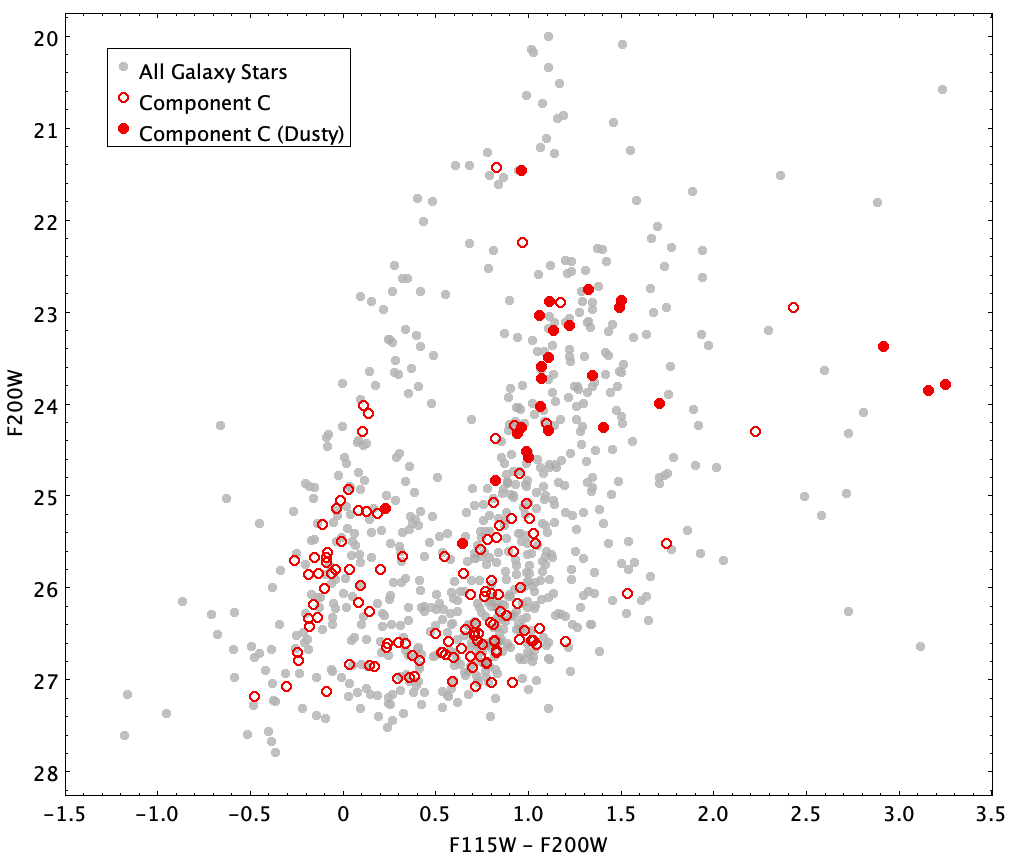}
\caption{
Comparison of the F200W vs.\ F115W--F200W CMD for stellar sources extracted from Component C (open red circles) against those of I~Zw~18's main body (gray dots).
With F200W vs.\ F115W--F200W ($n$ = 145), we are able to recognize the similarities in stellar content between the two systems.
A lack of detections on the brighter end of the left branch suggests a dearth of young, hot UMS stars in Component C, while dusty source candidates (red filled circles) appear to predominantly inhabit the bright end of the evolved star branch.
}
\label{fig:Component_C_CMD}
\end{figure*}

\begin{deluxetable}{lc} 
\label{tab:pointsourcedetections}
\tabletypesize{\small}
\tablewidth{0pt}
\tablecaption{
I~Zw~18 Point-source Summary
}
\tablehead{
\colhead{Source Type}&\colhead{No.\ of Sources}
}
\startdata
Young stars branch & 116 \\
$\rightarrow$ Dust-enshrouded & 20 \\
Evolved stars branch & 226 \\
$\rightarrow$ Dust-enshrouded & 99 \\
\hline
YSO candidates & 15 \\
\enddata
\end{deluxetable}

\subsection{Component C} 
\label{sec:Component_C}

The companion system to I~Zw~18, Component C, is offset from the main galaxy by $\sim$36\arcsec.
We present a comparison of the F200W vs.\ F115W--F200W CMD with Component C point sources (open red circles) overlaid upon stars found in the main body (gray dots) as Figure \ref{fig:Component_C_CMD}.

Contrasting the distributions of sources, we see stellar populations which are roughly self-consistent between I~Zw~18 and Component C ($n$ = 145), forming distinctive left- and right-side branches that represent young and evolved stars, respectively.
Of particular note is a comparative lack of Component C sources populating the bright end of the left-side branch, where the youngest UMS stars are expected to be located.
This suggests that Component C is home to a generally older population of stars, where the youngest and hottest UMS sources have already transitioned to later evolutionary stages.
An investigation of dusty targets, including selecting objects above the F200W vs.\ F115W--F200W CMD TRGB (F200W $\approx$ 25.6) and with F115W--F444W color $>$ 0.5, yields a small number ($n$ = 26) of sources denoted as filled red circles in Figure \ref{fig:Component_C_CMD}.
These appear to predominantly inhabit the bright end of the F200W vs.\ F115W--F200W CMD's evolved stars branch.
See \citet{bib:Bortolini2024} for an exploration of the SFH for Component C.


\section{Summary} 
\label{sec:summary}

We present for the first time JWST near- and mid-IR imaging data using NIRCam and MIRI of the XMP SF dwarf galaxy I~Zw~18.
The extremely low levels of heavy-element enrichment and high level of active star formation make it an excellent accessible analog for the types of small systems which are expected to have been ubiquitous in the very early Universe, and acted as major contributors to its overall chemical enrichment and reionization.
With eight wide-band filters combining NIRCam (F115W, F200W, F356W, and F444W) and MIRI (F770W, F1000W, F1500W, and F1800W), we show deep, high-angular-resolution images at IR wavelengths comparable to that of optical-band HST data, and improved by over an order of magnitude over that of existing mid-IR Spitzer images.
Our NIRCam data clearly show extreme populations of bright, recently-formed massive stars, located predominantly among two central lobes of star formation (the NW and SE components).
In addition, our MIRI imaging data demonstrate the bulk mid-IR emission characteristics of I~Zw~18 at previously unattainable resolution, exhibiting a predominantly wispy and clumpy nature in the NW and SE components, respectively.

Utilizing the four NIRCam filters, we have constructed CMDs and performed color-cut analyses from which we identified candidate populations of dusty evolved stars (RSGs and AGB stars) in this galaxy, alongside bright YSOs.
We employ a CMD utilizing the shortest-wavelength NIRCam filters (F200W vs.\ F115W--F200W), which demonstrate the greatest point-source sensitivity and therefore possess the largest number of sources, to analyze these populations.
Having estimated the location of the TRGB, we divide those sources which lie above it (F200W $\approx$ 25.6) into left- and right-branch populations, pertaining to younger (UMS) and older (evolved) stars, respectively.
Harnessing next a CMD with wider color baseline (F444W vs.\ F115W--F444W), we identify subsets of dust-enshrouded source candidates as those found above the TRGB and with color values $>$ 0.5, many of which are expected to be C-rich TP-AGB stars, but which also include O-rich AGB stars, RSGs, and bright YSOs.

Further investigation of YSO or YSO cluster candidates was accomplished via the F200W--F444W vs.\ F115W--F200W CCD, building upon techniques implemented for recent JWST study of the NGC 346 SF region in the SMC.
A small population of unresolved sources which satisfy the criteria F115W--F200W $>$ 1.2 and F200W--F444W $>$ 1.0 were isolated, with their locations in the galaxy correlated with MIRI F770W emission in the SE and NW lobes of star formation.
Our findings suggest that the prevailing demographics of I~Zw~18's NW and SE star-formation regions reflect that of younger and older stellar populations, respectively, consistent with a staggered SFH.

Point sources affiliated with I~Zw~18's companion system, Component C, were also extracted.
A similar analysis was performed, finding CMD structures closely matching those constructed from the galaxy's main body.
We find a comparative lack of detections in the CMD location representative of Component C's young, hot star population, suggesting that the bulk of its most recent star-formation episode took place longer ago than that of I~Zw~18's main body, consistent with the literature.
In addition, the small number of detected dust-enshrouded source candidates are preferentially located in the bright end of the evolved stars branch.

\section{Acknowledgements}
\begin{acknowledgments}

This work is based on observations made with the NASA/ESA/CSA James Webb Space Telescope. The data were obtained from the Mikulski Archive for Space Telescopes (MAST) at the Space Telescope Science Institute, which is operated by the Association of Universities for Research in Astronomy, Inc., under NASA contract NAS 5-03127 for JWST.
These observations are associated with program No.\ 1233.
The specific observations analyzed can be accessed via DOI:10.17909/3c1d-6182.

A.S.H. is supported in part by an STScI Postdoctoral Fellowship.
A.S.H. extends thanks to Steven R. Goldman for his assistance in clarifying details pertaining to dust production by AGB stars.
L.L. acknowledges support from the NSF through grant 2054178.
O.C.J. acknowledges support from an STFC Webb fellowship. 
C.N. acknowledges the support of an STFC studentship.
M.M. and N.H. acknowledge that a portion of their research was carried out at the Jet Propulsion Laboratory, California Institute of Technology, under a contract with the National Aeronautics and Space Administration (grant No.\ 80NM0018D0004).
M.M. and N.H. acknowledge support through NASA/JWST grant No.\ 80NSSC22K0025.
K.J. acknowledges support from the Swedish National Space Agency.
O.N. was supported by the STScI Postdoctoral Fellowship, and the NASA Postdoctoral Program at NASA Goddard Space Flight Center, administered by Oak Ridge Associated Universities under contract with NASA.
G.S.W. acknowledges support of UKSA and STFC.
J.B. and P.R. thank the Belgian Federal Science Policy Office (BELSPO) for the provision of financial support in the framework of the PRODEX Programme of the European Space Agency (ESA).
NIRCam and MIRI images created using the \texttt{jdaviz} tool \citep{bib:jdaviz}.

\emph{Facility:} JWST (MIRI, NIRCam).

\emph{Software:} astropy \citep{bib:Astropy2013, bib:Astropy2018, bib:Astropy2022}, \texttt{image1overf.py} \citep{bib:Willott2022}, jdaviz \citep{bib:jdaviz}, JHAT \citep{bib:Rest2023}, \textsc{starbugii} \citep{bib:Starbug}, \textsc{topcat} \citep{bib:Taylor2005}.

\end{acknowledgments}


\bibliography{main}{}
\bibliographystyle{aasjournal}

\end{document}